\documentclass[prb, aps, twocolumn, asmath, amssymb, floatfix, superscriptaddress,longbibliography]{revtex4-2}
\usepackage[dvips]{graphics}
\usepackage{adjustbox}
\usepackage{color}
\usepackage{float}
\usepackage{natbib}
\usepackage[dvipsnames]{xcolor}
\setcitestyle{numbers}
\setcitestyle{square}
\definecolor{dred}{rgb}{0,0,0.6}
\definecolor{NavyBlue}{rgb}{0, 0, 128}
\definecolor{RoyalBlue}{rgb}{0.255,0.41,0.884}
\usepackage{graphicx}
\usepackage{bm}
\usepackage{amssymb}
\usepackage{amsmath}
\usepackage{mathtools}
\usepackage{dcolumn}   
\usepackage[colorlinks, linkcolor=NavyBlue,citecolor=NavyBlue,urlcolor=NavyBlue]{hyperref}
\usepackage[all]{hypcap} 
\usepackage{physics}
\usepackage{enumerate}
\usepackage{braket}        
\usepackage{overpic}
\usepackage{amsfonts}
\usepackage{dsfont}
\usepackage[mathscr]{eucal}
\usepackage{hyperref}
\usepackage{setspace}
\usepackage{url}

\newcommand{\mb}{\mathbf}

\newcommand{\prbsubref}[2]{\hyperref[#1]{\ref*{#1}{(\subref*{#2})}}}

\usepackage[normalem]{ulem}
\definecolor{lime}{HTML}{A6CE39}
\usepackage{sidecap,tikz}
\DeclareRobustCommand{\orcidicon}{\hspace{-1.0mm}
	\begin{tikzpicture}
		\draw[lime, fill=lime] (0.0,0.0) 
		circle [radius=0.15] 
		node[white] {{\fontfamily{qag}\selectfont \tiny \,ID}};
		\draw[white, fill=white] (-0.0525,0.095) 
		circle [radius=0.007];
	\end{tikzpicture}
	\hspace{-3.0mm}
}
\foreach \x in {A, ..., Z}{\expandafter\xdef\csname orcid\x\endcsname{\noexpand\href{https://orcid.org/\csname orcidauthor\x\endcsname}
		{\noexpand\orcidicon}}
}

\begin{document}
\newcommand{\Z}{\mathbb{Z}}

\title{Topological Magnons and Giant Orbital Nernst Effect in a Zigzag Kitaev Antiferromagnet}

\author{Shreya Debnath$^\S$\orcidA{}}
\email[]{d.shreya@iitg.ac.in} 

\author{Saurabh Basu}
\email[]{saurabh@iitg.ac.in}
\affiliation{Department of Physics, Indian Institute of Technology Guwahati, Guwahati-781039, Assam, India}

\begin{abstract}
The exploration of topological and transport properties of collinear antiferromagnets and the role of Kitaev interactions in realising topological states therein have rarely been systematically addressed in literature. In this context, we consider a zigzag-ordered antiferromagnet with both extended Kitaev and Dzyaloshinskii-Moriya interactions (DMI) in presence of an external magnetic field to focus on the topological phases demonstrated by the magnon band structure and validated by the transport properties. The hybridization between the up- and down-spin sectors carries evidences of opening bulk gaps in the magnon band structure, giving rise to nontrivial topological phases characterized by finite Chern numbers, chiral edge modes, and a nonzero thermal Hall conductivity. Furthermore, generally speaking, a finite magnon orbital moment can exist and contribute to the Nernst response even when the net spin moment vanishes owing to the fundamental independence of the spin and orbital magnetizations. This motivates us to investigate the magnon orbital moment, orbital Berry curvature, and the resulting orbital Nernst conductivity associated with the magnon bands. We find that a giant orbital Nernst conductivity emerges even in the absence of an external magnetic field. Moreover, the distinction between different topological phases is more lucidly manifested via the orbital Nernst conductivity, thereby highlighting an enhanced sensitivity of the orbital transport to the underlying band topology. For completeness, we briefly discuss the scenario corresponding to a N\'eel-ordered spin alignment, which leads to a vanishing Chern number and consequently suppressed thermal Hall and orbital Nernst conductivities compared to the zigzag-ordered case, even in the presence of DMI and Kitaev interactions.
\end{abstract}
   
\maketitle


\section{INTRODUCTION}
\label{Introduction}
In the era of modern quantum devices, topological spin transport in two-dimensional magnetically ordered systems has attracted significant attention. Analogous to electronic systems, magnons, which are collective excitations of spins, have become a central focus of research not only in magnetically ordered materials, but also in frustrated magnets~\cite{Anderson1973,Lhuillier2005,Moessner2006}, quantum spin liquids~\cite{Balents2010,Zhou2017, Savary2017,Norman2016,Broholm2020}, and skyrmionic systems~\cite{Pan2024, Maeland2023}. As charge-neutral bosonic quasiparticles, magnons enable transport features without Joule heating and have emerged as promising platforms for realising nontrivial topological states.
Similar to the more widely studied electronic band topology, magnon band topology has also been explored in various magnetically ordered systems, including ferromagnets~\cite{magnon4_2016, magnon6_2021, PhysRevB.104.144422, li2021magnonic, moulsdale2019unconventional, anomalousTHE1, honeycomb1, PhysRevB.85.134411, PhysRevX.11.021061}, collinear~\cite{Neumann2022} and canted antiferromagnets~\cite{PhysRevB.98.094419, Owerre2017, Debnath2025II}, and altermagnets~\cite{Khatua2025}. In particular, antiferromagnets offer several advantages over ferromagnets owing to their strong exchange interactions and staggered spin ordering with zero net magnetisation. These properties provide enhanced robustness against external and stray magnetic fields and support ultrafast spin dynamics in the terahertz frequency regime~\cite{Kleinherbers2024}. 
In literature, magnon-based spin transport has primarily been investigated through phenomena such as the thermal Hall effect (THE)~\cite{Murakami2017,Katsura2010,Onose2010,Ideue2012,Hirschberger2015,Malki2017,Owerre2017,Laurell2018,Akazawa2020,Dias2023,He2024,Debnath2024,Debnath2025,Lee2015,HirschbergerII2015}, the spin Seebeck effect~\cite{Seebeck}, and the spin Nernst effect~\cite{NernstEffect1, NernstEffect2, NernstEffect3}. More recently, the orbital magnetic moment of magnons demonstrated an additional and significant role in contributing to the Nernst response, which we explore in the subsequent discussion.

To put our discussion on the Nernst effect in perspective, we choose a lattice geometry that has made appealing consequences in condensed matter physics, especially discovery of graphene. Apart from studying fascinating electronic properties and topology, the underlying geometry, namely honeycomb lattice has proven to be an ideal platform for studying magnonic phenomena, both theoretically and experimentally, as evidenced by the discovery of magnetic materials such as $\text{FePS}_3$~\cite{Lee2016}, $\text{NiPS}_3$~\cite{Kuo2016}, and $\text{CrI}_3$~\cite{Huang2017}. Previous studies of magnon band topology have primarily focused on ferromagnets and noncollinear antiferromagnets with finite net magnetisation, where spin–orbit coupling (SOC) and broken inversion symmetry lead to band gaps via the DMI, analogous to the Haldane flux in a Chern insulator.
Further, in the context of nontrivial magnon band topology, extended Kitaev interaction plays a crucial role in addition to the DMI.  
More recently, it has been shown that collinear antiferromagnets with zero net magnetisation can also host nontrivial magnon topology under an external magnetic field, often arising from the hybridization between different quasiparticle sectors~\cite{To2023}. In this context, realising topological magnon states driven solely by hybridization between distinct spin sectors in a collinear antiferromagnet would be particularly compelling.

In honeycomb lattices, several types of collinear antiferromagnetic order, such as Néel, stripe, and zigzag phases may emerge owing to competing magnetic interactions, notably, the Heisenberg exchange interaction and magnetic anisotropy~\cite{Lee2018, To2023}. In this work, we primarily focus on the zigzag phase. There are plenty of materials that possess zigzag ordering in its ground state. Moreover, beyond the Heisenberg exchange, staggered spin ordering in honeycomb lattices is also strongly influenced by bond dependent Kitaev interaction. The resulting phase diagrams depicted various phases have been extensively studied for spin-$\tfrac{1}{2}$ systems, where Kitaev interaction stabilizes zigzag, stripe, and quantum spin liquid phases~\cite{Fouet2001, Chaloupka2010}. However, comparatively few realizations are known for spin $S\geq1$ systems~\cite{Chen2021II, Zhou2021, Badrtdinov2021}. Notably, $\text{Na}_2\text{Ni}_2\text{TeO}_6$ has been shown to exhibit zigzag order driven by extended Kitaev interactions, as revealed by neutron scattering experiments~\cite{Samarakoon2021}. More recently, $\text{KNiAsO}_4$ has also been proposed as a zigzag-ordered system with a band spectrum consistent with extended Kitaev interactions~\cite{Taddei2023}.

\begin{figure}[t]
    \centering
    \includegraphics[width=1\linewidth]{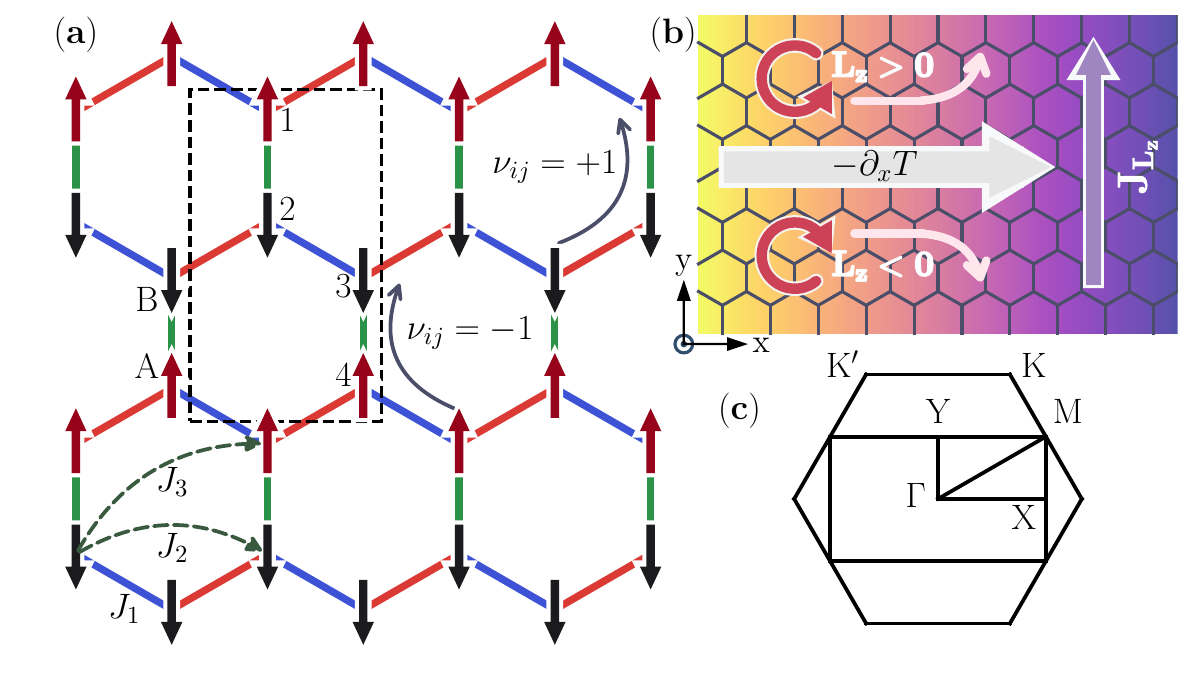}
    \caption{(a) Schematic diagram of a zigzag-ordered honeycomb antiferromagnet. The red and black arrows represent the up and down spins, respectively. $J_m$ ($m = 1,2,3$) denote the strengths of the Heisenberg exchange interactions. The dotted rectangular region indicates the magnetic unit cell, which consists of four lattice sites. The factor $\nu_{ij}$ takes values of $\pm 1$ depending on the ordering of the interacting spins. (b) Schematic diagram of the magnon orbital moment profile. $\partial_x T$ denotes a temperature gradient applied along the $x$-direction. $L_z$ represents the orbital moment, which takes positive(negative) values for right(left)-circularly propagating magnons. $J_{L_z}$ denotes the net transverse orbital current flowing along the $y$-direction. (c) The full hexagon corresponds to the original crystallographic Brillouin zone, whereas the rectangular region corresponds to the magnetic Brillouin zone.}
    \label{Model_zigzag}
\end{figure}
In the context of magnon band spectrum, a limited number of studies have investigated zigzag-ordered antiferromagnets in presence of DMI~\cite{To2024} and magnon–phonon coupling~\cite{To2023}. The former primarily focuses on electronic polarization induced by the magnon Nernst effect, while the latter reports the emergence of topological magnon–polaron states accompanied by a magnon Hall effect and a spin Nernst effect. Taking a clue from these results, in this paper we consider a zigzag-ordered honeycomb antiferromagnet as shown in Fig.~\ref{Model_zigzag}(a) that includes both DMI and extended Kitaev interactions. 
An extended Kitaev interaction consists of bond-dependent Ising-like interactions between neighbouring spins, along with an additional bond-dependent off-diagonal exchange interaction. Similar to the DMI, it originates from the presence of SOC~\cite{Kim2015}. In presence of both these interactions, we have shown how the hybridization between magnons originating from distinct spin sectors can induce topologically hybridized states and renders opening of a bulk gap, without involving other quasiparticle interactions. As a part of investigating transport features, we have studied thermal Hall conductivity. In addition, we have investigated how distinct topological phases can influence an intrinsic property of the magnons, namely, the magnon orbital moment. This orbital moment originates from the self-rotation of the magnon wave packets and serves as the magnonic analogue of the gauge-invariant electronic orbital moment, giving rise to a transverse magnon orbital current associated with magnon orbital Nernst effect (MONE)~\cite{Neumann2020}. Unlike THE, it is an intrinsic property and hence does not require breaking of the time-reversal symmetry (TRS). Additionaly, unlike spin Nernst conductivity, orbital Nernst conductivtiy may have finite values even in the absence of SOC~\cite{To2024}. In Fig.~\ref{Model_zigzag}(b), we present a schematic diagram illustrating the MONE, where in presence of a thermal gradiant, the left- and right-circular magnon wave packets propagate in opposite directions~\cite{An2025}. This counter propagating packets lead to orbital moment accumulation at the edges and gives rise to a transverse orbital current. In presence of both the DMI and the Kitaev interaction, the system behaves as a magnetic insulator, and the interplay between these interactions, as we shall see, leads to interesting outcomes in the context of the THE and MONE.

The paper is organized as follows. In Sec.~\ref{Formalism_zigzag}, we construct the model Hamiltonian for a zigzag-ordered Kitaev antiferromagnet and discuss the relevant symmetries in presence of different magnetic interactions. In Sec.~\ref{spectral_properties_zigzag}, we present the magnon bulk band structure in the presence of DMI and extended Kitaev interactions and analyze the topological properties of these bands via studying the Berry curvature, Chern numbers and investigating the phase transitions occuring therein. Sec.~\ref{edgemodes_zigzag} is devoted to the discussion of bulk–boundary correspondence and the emergence of chiral edge states.
In Sec.~\ref{Orbital_Berry_curvature_zigzag}, we introduce the magnon orbital moment and the magnon orbital Berry curvature and examine their behaviour for all the relevant bands vis-a-vis all the topological phases. The thermal Hall conductivity and the orbital Nernst conductivity are discussed in Sec.~\ref{Transport_zigzag}. In Sec.~\ref{Neel_antiferromagent}, we provide a brief discussion of the Néel-ordered antiferromagnet in the presence of both DMI and Kitaev interactions. Finally, we summarize and highlight the main findings of this work in Sec.~\ref{conclusion_zigzag}.

\begingroup
\allowdisplaybreaks
\section{Model Hamiltonian and Symmetries}
\label{Formalism_zigzag}
As stated earlier, the collinear zigzag ordering in the honeycomb structure requires several magnetic interactions so that it can be regarded as an energy-minimum state associated with spin fluctuations. Hence, to begin with, we write the spin Hamiltonian as,
\begin{equation}
    \begin{split}
       H &= J_m \sum_{\langle i,j \rangle_m} \mathbf{S}_i \cdot \mathbf{S}_j
        + D \sum_{\langle\!\langle i,j \rangle\!\rangle} \nu_{ij} \hat{z} \cdot (\mathbf{S}_i \times \mathbf{S}_j) \\
        &~~~~~ - \mu_B B_0 \sum_i S_i^z 
        - \Delta \sum_i (S_i^z)^2 \\
        & + \sum_{\langle i,j \rangle^\gamma} \Big[
        \mathcal{K} S_i^\gamma S_j^\gamma
       + \Gamma_{\scriptscriptstyle \mathcal{K}} (S_i^\alpha S_j^\beta + S_i^\beta S_j^\alpha) \\
       &  + \Gamma_{\scriptscriptstyle \mathcal{K}}^{\prime} (S_i^\alpha S_j^\gamma + S_i^\gamma S_j^\alpha
       + S_i^\beta S_j^\gamma + S_i^\gamma S_j^\beta)
   \Big],
    \end{split} \label{Model_Hamiltonain_zigzag}  
\end{equation}
where $J_m$ ($m=1,2,3$) are the Heisenberg exchange interaction strengths corresponding to the first, second, and third nearest neighbouring sites on a honeycomb lattice. $D$ is the coefficient of the DMI acting between next-nearest-neighbour sites, and $\nu_{ij} = \pm 1$ corresponds to the bond orientation being anticlockwise or clockwise, respectively as shown in Fig.~\ref{Model_zigzag}(a). Further, the third and the fourth terms correspond to the Zeeman and anisotropy energies, respectively, associated with each site. Here, $B_0$ denotes the magnetic field, and $\Delta$ represents the anisotropy strength. 
Additionally, the last term in Eq.~\eqref{Model_Hamiltonain_zigzag} represents the bond-dependent extended Kitaev interaction acting between neighbouring sites, where $\mathcal{K}$, $\Gamma_{\scriptscriptstyle \mathcal{K}}$, and $\Gamma_{\scriptscriptstyle \mathcal{K}}^{\prime}$ are the corresponding interaction strengths. Here, $\gamma$ denotes the bond type, and depending on its value, the indices $\alpha$ and $\beta$ are distinct and satisfy $\alpha \neq \beta \neq \gamma$. Furthermore, the lattice vectors connecting the first-, second-, and third-nearest neighbours are denoted by $\boldsymbol{\delta}_i$, $\boldsymbol{\gamma}_i$, and $\boldsymbol{\xi}_i$, respectively, and can be expressed as
\begin{equation}
    \begin{split}
        \boldsymbol{\delta}_1 &=(0,1)a_0,~\boldsymbol{\delta}_2 = (\frac{\sqrt{3}}{2}, -\frac{1}{2})a_0,~\boldsymbol{\delta}_3 = (-\frac{\sqrt{3}}{2}, -\frac{1}{2})a_0,\\
        \boldsymbol{\gamma}_1 &=(-\sqrt{3},0)a_0,~\boldsymbol{\gamma}_2 = (\frac{\sqrt{3}}{2}, \frac{3}{2})a_0,~\boldsymbol{\gamma}_3 = (\frac{\sqrt{3}}{2}, -\frac{3}{2})a_0,\\
        \boldsymbol{\xi}_1 &=(0,2)a_0,~\boldsymbol{\xi}_2 =(\sqrt{3},1)a_0,~\boldsymbol{\xi}_3 = (-\sqrt{3},1)a_0,
        \end{split}
\end{equation}
where $a_0$ is the lattice constant (set to unity in our calculations). 
\endgroup

To study spin fluctuations or low-energy magnon excitations in the one-magnon regime, we employ the linearised Holstein--Primakoff transformation to establish a correspondence between spin and bosonic operators. For the up ($A$-type sublattices) and down ($B$-type sublattices) spins associated with this collinear antiferromagnet, the transformations can be expressed as
\begin{subequations}\label{HP_zigzag}
\begin{align}
S_{A_i}^+ &= \sqrt{2S}\,a_i,
S_{A_i}^- = \sqrt{2S}\,a_i^\dagger,
S_{A_i}^z = S - a_i^\dagger a_i, \label{HP_zigzag_A}\\
S_{B_j}^+ &= \sqrt{2S}\,b_j^\dagger,
S_{B_j}^- = \sqrt{2S}\,b_j,
S_{B_j}^z = -S + b_j^\dagger b_j. \label{HP_zigzag_B}
\end{align}
\end{subequations}
where $S_i^+ = S_i^x + iS_i^y$ and $S_i^- = S_i^x - iS_i^y$, while $a_i^\dagger$($a_i$) and $b_j^\dagger$($b_j$) denote the creation(annihilation) operators for the $A$- and $B$-type sublattices, respectively. Moreover, in the zigzag ordered phase, the unit cell contains four nonequivalent sublattices, $A_1$, $B_2$, $B_3$, and $A_4$, as shown in Fig.~\ref{Model_zigzag}. The sublattices $A_1$ and $A_4$ undergo the up–spin transformation (that is, Eq.~\ref{HP_zigzag_A}), whereas $B_2$ and $B_3$ undergo the down–spin transformation (Eq.~\ref{HP_zigzag_A}), as specified in Eq.~\eqref{HP_zigzag}. Consequently, upon applying these transformations, the Hamiltonian takes the form given in Eq.~\eqref{real_H_zigzag} (See Appendix~\ref{spin_Hamiltonian_zigzag}).

Now, in the zigzag ordered phase, the Fourier-transformed Hamiltonian can be expressed in the Bogoliubov–de Gennes (BdG) form using the basis $(a_{1k}, b_{3,-k}^\dagger, a_{4k}, b_{2,-k}^\dagger, a_{1,-k}^\dagger, b_{3,k}, a_{4,-k}^\dagger, b_{2,k})^T$, and is given by,
\begin{equation}
    \mathcal{H}(\mb{k})=\begin{bmatrix}
                            A(\mb{k}) & B(\mb{k})\\
                            B^*(-\mb{k}) & A^*(-\mb{k})
                        \end{bmatrix},\label{H_kspace_zigzag}
\end{equation}
where the submatrices $A(\mb{k})$ and $B(\mb{k})$ can be expressed as
\begin{subequations}
\begin{align}
    A(\mb{k})&=\begin{pmatrix}
                M_{A_1}(\mb{k}) & f_1(\mb{k}) & f_2(\mb{k}) & f_3(\mb{k})\\
                f_1^*(\mb{k}) & M_{B_3}(\mb{k}) & f_3(\mb{k}) & f_2(\mb{k})\\
                f_2^*(\mb{k}) & f_3^*(\mb{k}) & M_{A_4}(\mb{k}) & f_1(-\mb{k})\\
                f_3^*(\mb{k}) & f_2^*(\mb{k}) & f_1^*(-\mb{k}) & M_{B_2}(\mb{k})
    \end{pmatrix},\\
    B(\mb{k})&= \begin{pmatrix}
                 0 & 0 & g_1(\mb{k}) &  g_2(\mb{k})\\
                 0 & 0 & g_2(\mb{k}) &  g_1(\mb{k})\\
                 g_1(-\mb{k}) &  g_2(-\mb{k}) & 0 & 0\\
                 g_2(-\mb{k}) &  g_1(-\mb{k}) & 0 & 0
    \end{pmatrix},      
\end{align}
\end{subequations}
where all the components of the $A(\mb{k})$ and $B(\mb{k})$ matrices are given as
\begin{subequations}
\begingroup
\allowdisplaybreaks
\begin{align}
    \begin{split}
       &M(\mathbf{k}) = S(-J_1 + 2J_2(1 + \cos(\mathbf{k}\cdot\boldsymbol{\gamma}_1)) + 3J_3 + K),\\
       &M_{A_1}(\mathbf{k}) = M(\mathbf{k}) + \mu_B B_0 - 2DS\sin(\mathbf{k}\cdot\boldsymbol{\gamma}_1),\\
       &M_{B_2}(\mathbf{k}) = M(\mathbf{k}) - \mu_B B_0 + 2DS\sin(\mathbf{k}\cdot\boldsymbol{\gamma}_1),\\
       &M_{B_3}(\mathbf{k}) = M(\mathbf{k}) - \mu_B B_0 - 2DS\sin(\mathbf{k}\cdot\boldsymbol{\gamma}_1),\\
       &M_{A_4}(\mathbf{k}) = M(\mathbf{k}) + \mu_B B_0 + 2DS\sin(\mathbf{k}\cdot\boldsymbol{\gamma}_1),
\end{split}\\
\begin{split}
        &f_1(\mb{k}) = 2J_2 S(\cos{(\mb{k} \cdot \boldsymbol{\gamma}_2)} + \cos{(\mb{k} \cdot \boldsymbol{\gamma}_3)})\\
       &~~~~~~~-2DS(\sin{(\mb{k} \cdot \boldsymbol{\gamma}_2)} + \sin{(\mb{k} \cdot \boldsymbol{\gamma}_3)}),\\
       &f_2(\mb{k}) = J_1 S (e^{i\mb{k} \cdot \boldsymbol{\delta}_2} + e^{i\mb{k} \cdot \boldsymbol{\delta}_3})
       +\frac{\mathcal{K}S}{2}(e^{i\mb{k} \cdot \boldsymbol{\delta}_2} + e^{i\mb{k} \cdot \boldsymbol{\delta}_3}),\\
       &f_3(\mb{k}) = J_1S e^{i\mb{k} \cdot \boldsymbol{\delta}_1} + J_3 S (e^{-i\mb{k} \cdot \boldsymbol{\xi}_1}+ e^{i\mb{k} \cdot \boldsymbol{\xi}_2} + e^{i\mb{k} \cdot \boldsymbol{\xi}_3}),
\end{split}\\
\begin{split}
       &g_1(\mb{k})=\frac{\mathcal{K}S}{2}(e^{i\mb{k} \cdot \boldsymbol{\delta}_3} - e^{i\mb{k} \cdot \boldsymbol{\delta}_2}) + i \Gamma_{\scriptscriptstyle \mathcal{K}}^{\prime} S (e^{i\mb{k} \cdot \boldsymbol{\delta}_3} + e^{i\mb{k} \cdot \boldsymbol{\delta}_2}),\\
        &g_2(\mb{k})= i \Gamma_{\scriptscriptstyle \mathcal{K}} S e^{i\mb{k} \cdot \boldsymbol{\delta}_1}.
    \end{split} 
\end{align}
\endgroup
\end{subequations}
Here, $\mathcal{H}(\mb{k})$ satisfies a generalized eigenvalue problem. In this context, we introduce a non-Hermitian matrix, $\mathcal{H}^\prime(\mb{k})$ defined via $\mathcal{H}^\prime(\mb{k}) = \eta\mathcal{H}(\mb{k})$, where $\eta = \text{diag}(1,-1,1,-1,-1,1,-1,1)$. As long as we consider a stable ground-state configuration corresponding to a collinear antiferromagnetic zigzag order, the matrix $\mathcal{H}^\prime(\mb{k})$ yields real eigenspectra, despite its non-Hermitian form. Furthermore, since we are dealing with bosonic excitations in this transformed Hamiltonian, only the positive-energy branches have physical relevance.

Although the Hamiltonian no longer explicitly retains spin information after having the spin–boson transformation, the symmetries of the system can still be analysed through effective symmetry operators constructed according to the structure of the chosen magnon basis. In the absence of an external magnetic field ($B_0$), the system preserves an effective TRS, formulated as a combination of the conventional TRS operator, which reverses the spin direction and a lattice translation dictated by the underlying spin configuration, ensuring that the overall spin configuration remains the same. Similarly, in the magnon basis, the corresponding effective TRS operator is given by the paraunitary operator $\Theta = \mathcal{U}\mathcal{K}$, where $\mathcal{U} = \sigma_3 \otimes I_2 \otimes \sigma_3$ is the unitary part and $\mathcal{K}$ denotes complex conjugation. In the absence of $B_0$, $\Theta$ satisfies the symmetry condition, $\Theta \mathcal{H}(\mathbf{K}) \Theta^{-1} = \mathcal{H}(-\mathbf{K})$, but this relation no longer holds once the magnetic field turned on, indicating that the effective TRS is broken. However, aside from TRS, the zigzag-ordered phase also respects a glide-mirror symmetry $\mathcal{M}_y \tau$, where $\mathcal{M}_y$ denotes the mirror reflection about the plane perpendicular to the $y$-axis and $\tau$ corresponds to half lattice translation. Furthermore, this symmetry is represented by the unitary operator $\mathcal{U}_{\small{\mathcal{M}}} = I_2 \otimes \sigma_3 \otimes \sigma_3$, which satisfies $\mathcal{U}_{\small{\mathcal{M}}} \mathcal{H}(k_x, k_y)\mathcal{U}_{\small{\mathcal{M}}}^{-1} = \mathcal{H}(k_x, -k_y)$. While an external magnetic field does not break this symmetry, antisymmetric exchange interactions, such as, the DMI, or bond-dependent interactions, for example, the extended Kitaev interaction, do break the $\mathcal{M}_y \tau$ symmetry.

\section{Result and Discussion}
\label{Result_zigzag}
As discussed earlier, the zigzag-ordered state depends sensitively on long-range exchange interactions. Moreover, it is analytically challenging to determine the full parameter set for which this collinear zigzag configuration remains stable as the ground state. To address this issue, we adopt the initial parameter values reported for a representative material, $\text{FePS}_3$ ($J_1 =1.49~\text{meV},~J_2 = 0.04~\text{meV},~J_3 = -0.6~\text{meV},~\Delta = -3.6~\text{meV}$)~\cite{To2023}. Since one of our objectives is to examine the interplay between the antisymmetric DMI and the extended Kitaev interaction, we keep these parameter values flexible and will discuss the implications of their inclusion later.
\begin{figure}[t]
    \centering
    \includegraphics[width=1\linewidth]{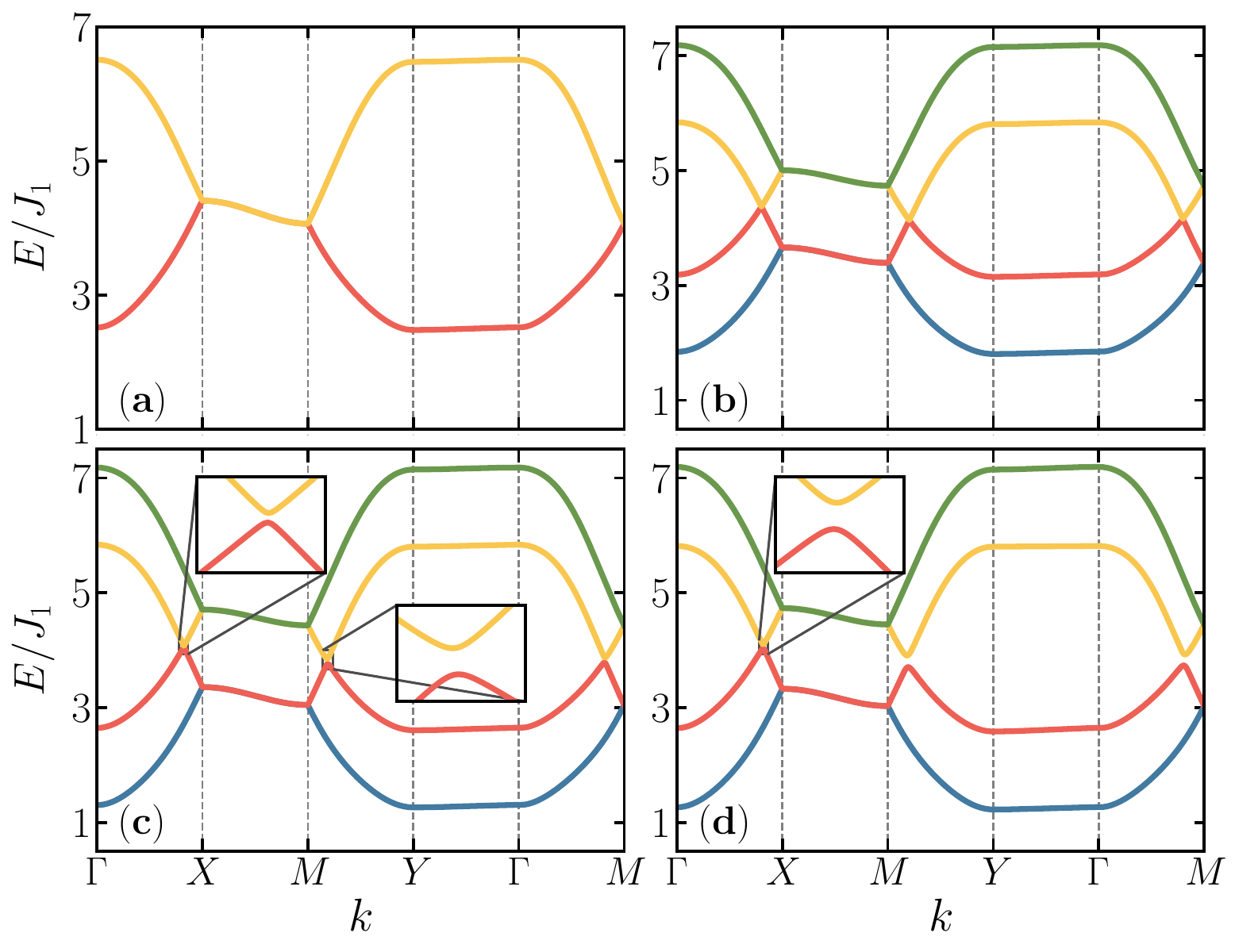}
    \caption{Bulk magnon band structures are shown along the high-symmetry path in the Brillouin zone.
            (a) Band structure in the absence of magnetic field, DMI, and Kitaev interaction, computed for parameter values
            $J_1 = 1.49~\text{meV}$, $J_2 = 
            0.04~\text{meV}$, $J_3 = -0.6~\text{meV}$, and $\Delta = -3.6~\text{meV}$.
            (b) Band structure after applying a magnetic field and DMI with $B_0 = 1~\text{meV}$ and $D = 0.3~\text{meV}$.
            (c) Further inclusion of the Kitaev interaction with $\mathcal{K} = 0.4~\text{meV}$.
            (d) Finally the band structure upon adding the extended Kitaev terms $\Gamma_{\scriptscriptstyle \mathcal{K}} = \Gamma_{\scriptscriptstyle \mathcal{K}}^\prime = 0.2~\text{meV}$ are presented.}
    \label{Zigzag_bandstructure}
\end{figure}
\subsection{Spectral properties and topological aspects of magnon bands}
\label{spectral_properties_zigzag}
All the symmetries discussed earlier can be directly inferred from the band structures. Owing to the zigzag ordering of the spin structure, the two-dimensional spin configuration is described by a rectangular magnetic unit cell, which in turn gives rise to a rectangular magnetic Brillouin zone (MBZ) as shown in Fig.~\ref{Model_zigzag}(c). All the band structures presented here are plotted along the high-symmetry points of the MBZ, namely $\Gamma = (0,0) \to X = (\pi/\sqrt{3}, 0) \to M = (\pi/\sqrt{3}, \pi/3) \to Y = (0, \pi/3) \to \Gamma \to M$. In the absence of an external magnetic field, all excitations corresponding to up- and down-spin sectors must remain degenerate. Consequently, in Fig.~\ref{Zigzag_bandstructure}(a), each of the two bands is twofold-degenerate where the up- and down-spin excitations cannot be distinguished. The band separation arises from distinct glide mirror eigenvalues, which becomes degenerate at the MBZ boundary along the $X$–$M$ symmetry path. Upon applying an external magnetic field perpendicular to the spin space, the degeneracy between the up- and down-spin sectors is lifted. However, along the $\Gamma$–$X$, $M$–$Y$, and $\Gamma$–$M$ paths, certain crossings (degeneracies) persist where the middle two bands, associated with the up- and down-spin excitations, remain gapless. Moreover, although the presence of DMI breaks the glide mirror symmetry, these band crossings remain intact, as shown in Fig.~\ref{Zigzag_bandstructure}(b).
Now, when a Kitaev interaction is introduced, the hybridization between different spin sectors leads to a band inversion of the two  middle bands, resulting in the opening of a gap between them, as illustrated in Fig.~\ref{Zigzag_bandstructure}(c). Furthermore, the inclusion of extended Kitaev interactions, namely the off-diagonal exchange terms ($\Gamma_{\scriptscriptstyle \mathcal{K}}$ and $\Gamma_{\scriptscriptstyle \mathcal{K}}^\prime$), further broadens the gap along the symmetry paths as shown in Fig.~\ref{Zigzag_bandstructure}(d). Notably, even in the absence of DMI, a band gap can be opened solely in the presence of extended Kitaev interactions, that is, when all $\mathcal{K}$, $\Gamma_{\scriptscriptstyle \mathcal{K}}$, and $\Gamma_{\scriptscriptstyle \mathcal{K}}^\prime$ are included. In contrast, a nonzero $\mathcal{K}$ alone or nonzero $\Gamma_{\scriptscriptstyle \mathcal{K}}$ and $\Gamma_{\scriptscriptstyle \mathcal{K}}^\prime$ terms by themselves are insufficient to lift the degeneracies.

\begin{figure}[t]
    \centering
    \includegraphics[width=1\linewidth]{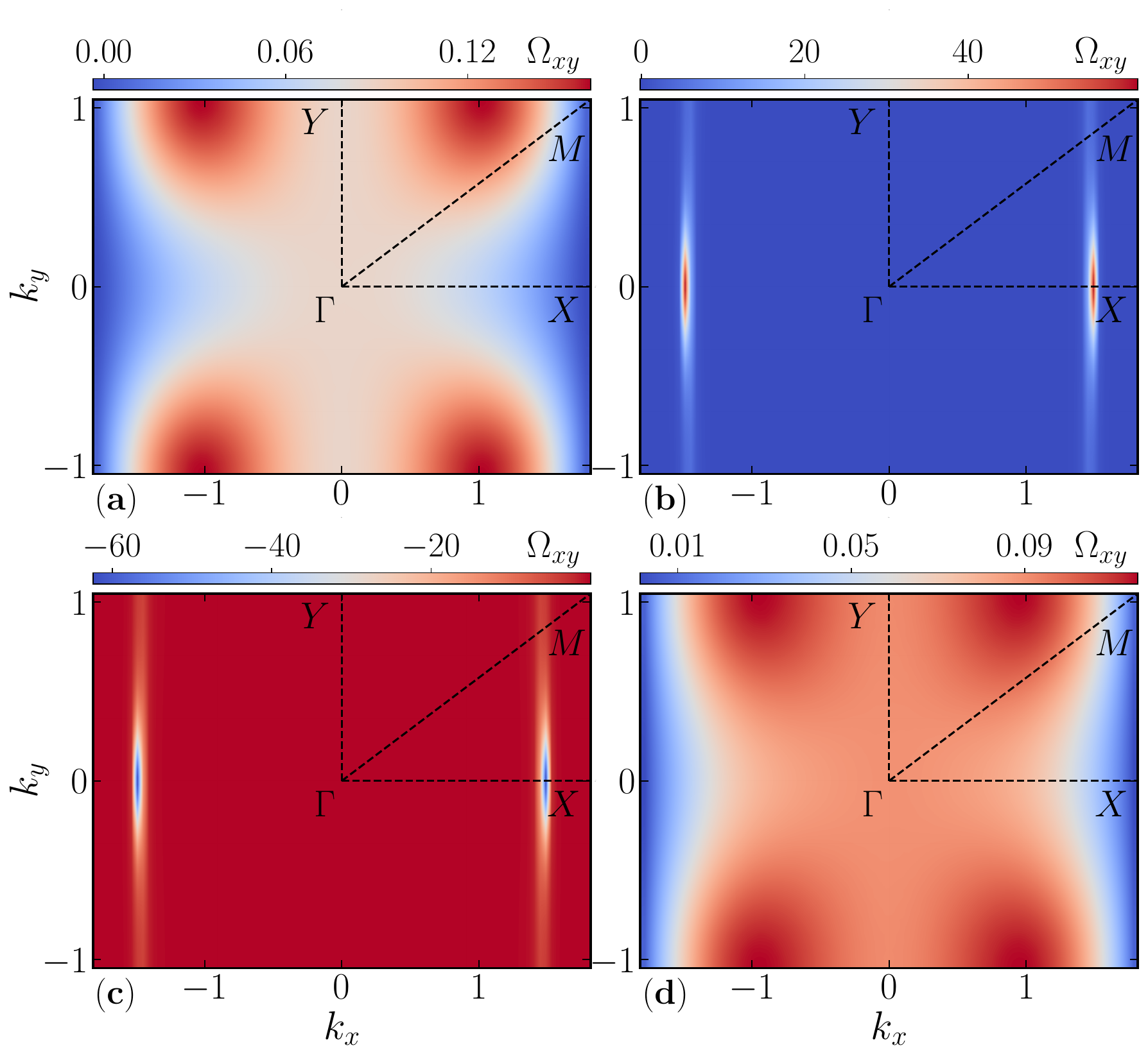}
    \caption{The Berry curvatures corresponding to (a) $n = 1$, (b) $n = 2$, (c) $n = 3$, and (d) $n = 4$ bands are shown for the parameter values $J_1 = 1.49~\text{meV}$, $J_2 = 0.04~\text{meV}$, $J_3 = -0.6~\text{meV}$, $\Delta = -3.6~\text{meV}$, $B_0 = 1~\text{meV}$, $D = 0.3~\text{meV}$, $\mathcal{K} = 0.4~\text{meV}$, and $\Gamma_{\scriptscriptstyle \mathcal{K}} = \Gamma_{\scriptscriptstyle \mathcal{K}}^\prime = 0.2~\text{meV}$.}
    \label{Berry_curvature_Zigzag}
\end{figure}

Since the effective TRS is already broken by the external magnetic field and the system undergoes hybridization between different spin sectors, the topological nature of the resulting band gaps can be examined by computing the Berry curvature over the MBZ, which is given by~\cite{Debnath2024}
\begin{equation}
    \Omega_{xy}^n(\mathbf{k}) = -2~\text{Im}\braket{\partial_{k_x} \psi_n(\mathbf{k})|\partial_{k_y} \psi_n(\mathbf{k})},\label{BC_zigzag}
\end{equation}
where, $n$ is the band index and $\ket{\psi_n(\mathbf{k})}$ is the corresponding wavefunction. Fig.~\ref{Berry_curvature_Zigzag} shows the Berry curvature corresponding to all the four ($n = 1,\cdots4$) bands. Although the band structure in Fig.~\ref{Zigzag_bandstructure} indicates that the $M$–$X$ path is symmetry-protected, leading to degeneracies between the pair of upper and lower bands, the Berry curvature remains well defined. By appropriately enumerating the energy bands, the corresponding eigenstates can be uniquely identified, enabling the Berry curvature of individual bands to be computed using Eq.~\eqref{BC_zigzag}.

Here, we observe that the dominant contribution to the nonzero Berry curvature originates from the two middle bands. For the lower ($n = 1$) and upper ($n = 4$) bands, a small yet finite positive Berry curvature is observed along the $M$–$Y$ symmetry path, whereas a negative Berry curvature emerges along the $M$–$X$ path, as shown in Figs.~\ref{Berry_curvature_Zigzag}(a),~(d). In contrast, along the $\Gamma$–$X$ path, the second band ($n =2$) exhibits a large positive Berry curvature, while the third band ($n = 3$) shows a correspondingly large negative Berry curvature, along with very large band curvatures, as illustrated in Fig.~\ref{Berry_curvature_Zigzag}.

One can find the Chern number of those corresponding bands given by,
\begin{equation}
    C_n = \frac{1}{2\pi}\int_{BZ}\Omega_{xy}^n (\mathbf{k}) d\mathbf{k}. \label{Chern_zigzag}
\end{equation}
However, due to the degeneracy along the $X$–$M$ path, the Chern number of an individual band may not be well defined. In such cases, we shall determine the combined Chern number of the two bands, which should yield an integer value corresponding to the topological phase.

For the parameter values used in Fig.~\ref{Zigzag_bandstructure}(c), namely $\mathcal{K} = 0.4~\text{meV}$ and $D = 0.3~\text{meV}$, the Chern number of the lower two bands is obtained as $-1$, while that of the upper two bands is $+1$. Upon introducing the off-diagonal interaction $\Gamma_{\scriptscriptstyle \mathcal{K}} = \Gamma_{\scriptscriptstyle \mathcal{K}}^\prime = 0.2~\text{meV}$, as shown in Fig.~\ref{Zigzag_bandstructure}(d), the Chern numbers swap to values $+1$ for the lower two bands and $-1$ for the upper two bands. This sign reversal indicates a topological phase transition, which must occur via a band-gap closing.

\begin{figure}[t]
    \centering
    \includegraphics[width=1\linewidth]{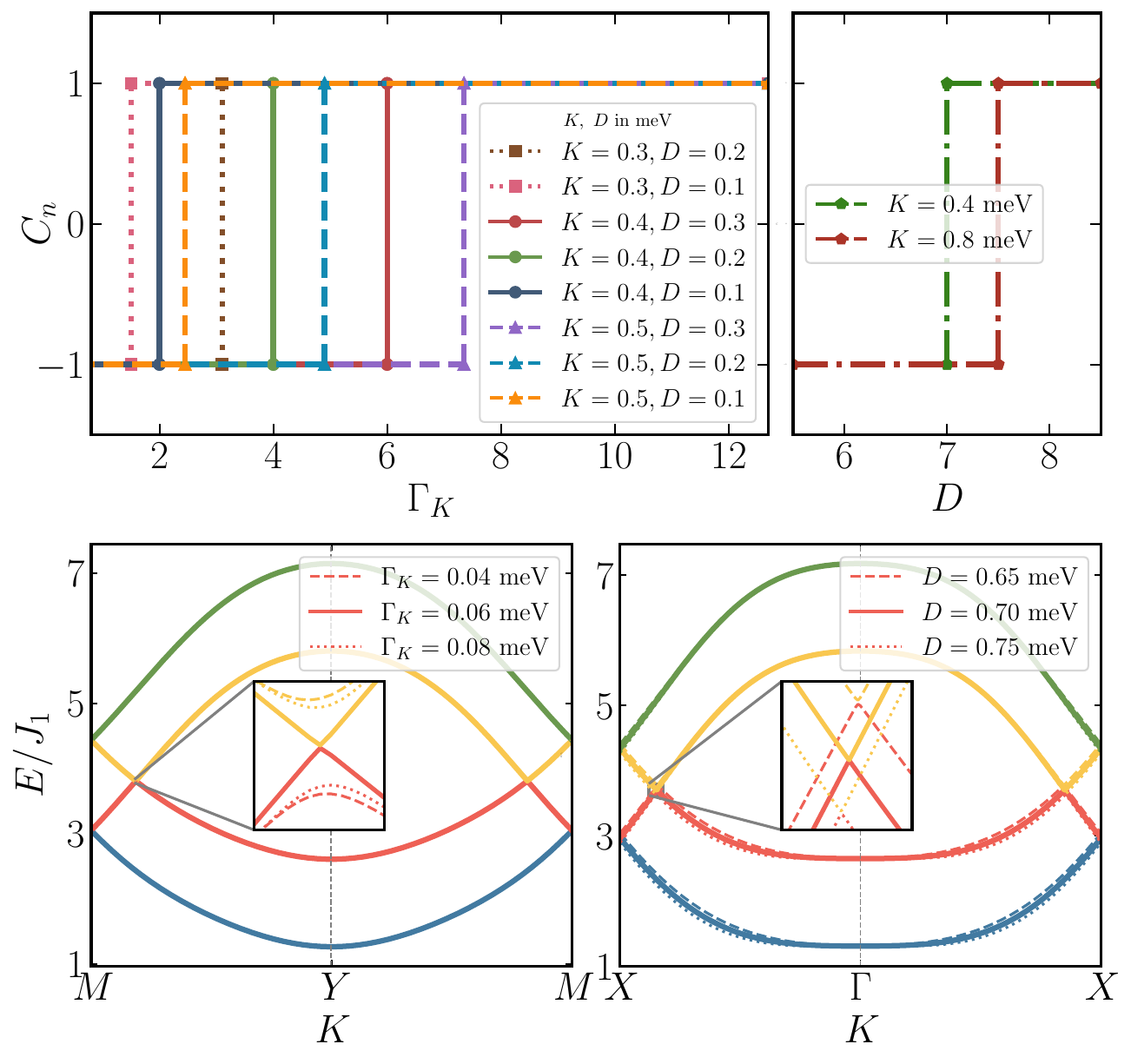}
    \caption{Phase transitions as a function of (a) $\Gamma_{\scriptscriptstyle \mathcal{K}}$ and (b) $D$ are demonstrated, where $C_n$ denotes the combined Chern number of the lower two bands. The gap-closing transitions obtained by varying (c) $\Gamma_{\scriptscriptstyle \mathcal{K}}$ and (d) $D$ are illustrated through the corresponding bulk band structures. Here, we consider $\Gamma_{\scriptscriptstyle \mathcal{K}} = \Gamma_{\scriptscriptstyle \mathcal{K}}^\prime$.}
    \label{phase_zigzag}
\end{figure}
To characterize this topological phase transition, we study the evolution of the Chern numbers as a functions of the interaction terms $\Gamma_{\scriptscriptstyle \mathcal{K}}$ (with $\Gamma_{\scriptscriptstyle \mathcal{K}}^\prime = \Gamma_{\scriptscriptstyle \mathcal{K}}$) and $D$, while keeping all other parameter values fixed, as shown in Fig.~\ref{phase_zigzag}. In Fig.~\ref{phase_zigzag}(a) keeping all the parameter fixed at $J_1 = 1.49~\text{meV}$, $J_2 = 0.04~\text{meV}$, $J_3 = -0.6~\text{meV}$, and $\Delta = -3.6~\text{meV}$, we have shown the total Chern number corresponding to the lower two bands. For larger values of $\mathcal{K}$, the phase transition occurs at smaller values of $\Gamma_{\scriptscriptstyle \mathcal{K}}$ for a fixed $D$. Conversely, for a fixed value of $\mathcal{K}$, reducing $D$ lowers the critical value of $\Gamma_{\scriptscriptstyle \mathcal{K}}$ at which the phase transition takes place, as shown in Fig.~\ref{phase_zigzag}(a). Furthermore, Fig.~\ref{phase_zigzag}(b) demonstrates that, in the absence of the off-diagonal interactions, increasing $\mathcal{K}$ shifts the occurrence of the phase transition to higher values of $D$.

Further, each of the phase transition occurs via a band-gap closing event, which can be directly observed in the band spectrum. In Fig.~\ref{phase_zigzag}(c), the band structure is shown for three representative values of $\Gamma_{\scriptscriptstyle \mathcal{K}}$ at $\mathcal{K} = 0.4~\text{meV}$ and $D = 0.3~\text{meV}$. Prior to the phase transition, the middle two bands (indicated by dashed lines) are gapped along the $M$–$Y$ symmetry path. At the critical value $\Gamma_{\scriptscriptstyle \mathcal{K}} = 0.06~\text{meV}$, the gap closes, as indicated by the solid lines, and subsequently reopens for $\Gamma_{\scriptscriptstyle \mathcal{K}} = 0.08~\text{meV}$, as shown by the dotted lines.

Similarly, in the absence of off-diagonal interactions, the gap-closing event associated with the topological phase transition occurs along the $X$–$\Gamma$ symmetry path, where the band gap closes at $D = 0.70~\text{meV}$. In this case, the band gap both before and after the gap-closing point remains relatively small compared to that shown in Fig.~\ref{phase_zigzag}(c). Moreover, as the strength of DMI is varied, the band extrema shift along the symmetry path.
\subsection{Magnonic edge modes in a ribbon}
\label{edgemodes_zigzag}
\begin{figure}[t]
    \centering
    \includegraphics[width=1\linewidth]{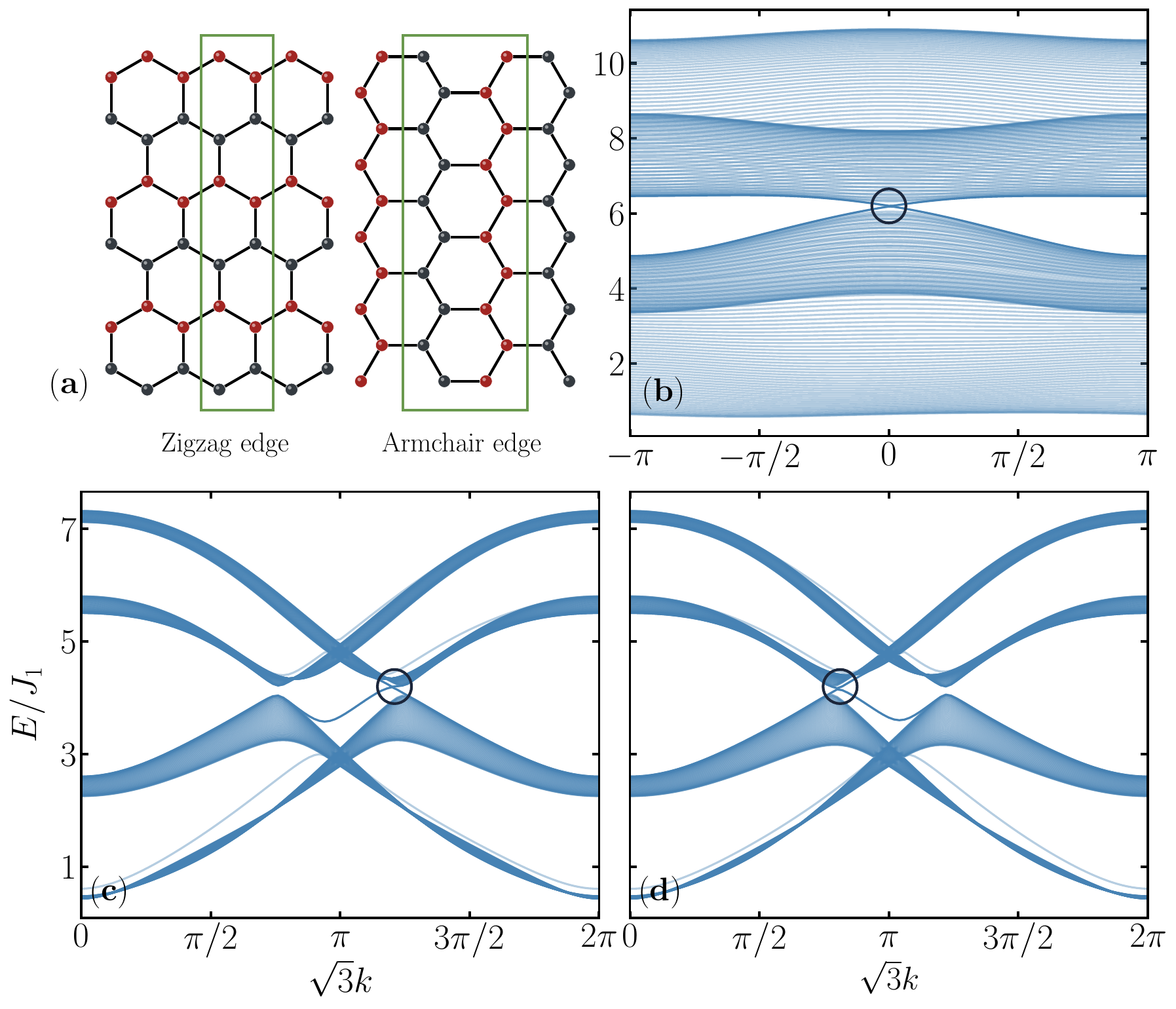}
    \caption{(a) Illustrations of the two distinct edge terminations in a semi-infinite zigzag antiferromagnet are shown. The bulk–boundary correspondence in (b) an armchair nanoribbon and (c) a zigzag nanoribbon is presented for the parameter values $J_1 = 1.49~\text{meV}$, $J_2 = 0.04~\text{meV}$, $J_3 = -0.6~\text{meV}$, $\Delta = -3.6~\text{meV}$, $B_0 = 1~\text{meV}$, $D = 0.3~\text{meV}$, $\mathcal{K} = 0.4~\text{meV}$, and $\Gamma_{\scriptscriptstyle \mathcal{K}} = \Gamma_{\scriptscriptstyle \mathcal{K}}^\prime = 0.8~\text{meV}$. (d) For another phase with $D = -0.3~\text{meV}$, the bulk–boundary correspondence is shown for a zigzag nanoribbon. The black circles highlight the crossings.}
    \label{edgestate_zigzag}
\end{figure}

So far, we have examined how, in the presence of DMI and extended Kitaev interactions, topological phases arise due to hybridization between two opposite spin sectors. These topological phases and the associated phase transitions have been identified through the bulk band structure along the high-symmetry paths within the MBZ, and by computing the Chern numbers corresponding to the gapped bands. However, it is also essential to substantiate these phases by analyzing the bulk–boundary correspondence. To this end, we consider a semi-infinite ribbon geometry that is finite along one direction and infinite along the other. For the zigzag-ordered antiferromagnetic system, two distinct ribbon geometries can be defined: one with zigzag edges and the other with armchair edges, as shown in Fig.~\ref{edgestate_zigzag}(a). Further, the number of crossings, among the edge modes depends on the value of the winding number, enumerated by the sum of the Chern numbers corresponding to the bands that lie below the gap and can be written as
\begin{equation}
    \mathcal{W}_i = \sum_{n = 0}^i C_n,
\end{equation}
where $n$ is the band index and $i$ denotes the gap index. The value of $\mathcal{W}_n$ denotes the number of crossing and its sign indicates the direction of propagation of the chiral edge current.

Here, depending on the Chern numbers of the bands, the magnitude of the winding number associated with the gapped region between the second and third bands is always equal to $1$. However, its sign depends on the specifics of the topological phase realized by the system. For instance, for $B = 1~\text{meV}$, $D = 0.3~\text{meV}$, $\mathcal{K} = 0.4~\text{meV}$, and $\Gamma_{\scriptscriptstyle \mathcal{K}} = \Gamma_{\scriptscriptstyle \mathcal{K}}^\prime = 0.8~\text{meV}$, the winding number corresponding to the middle (intervening between the lower and upper) gap ($i=2$) is $\mathcal{W}_2 = +1$. In this case, the edge current flows from left to right along the top edge and from right to left along the bottom edge of the ribbon. The resulting chiral edge-state crossings are shown for an armchair edge in Fig.~\ref{edgestate_zigzag}(b) and for zigzag edges in Fig.~\ref{edgestate_zigzag}(c). 
For an armchair edge, the edge-state crossing occurs at $\mathcal{K} = 0$, whereas for a zigzag edge, the crossing appears at $k > \pi$, indicating positive chirality. In contrast, for $D = -0.3~\text{meV}$, while keeping all other parameters same as before, the Chern number changes, resulting in $\mathcal{W}_2 = -1$. Consequently, for a zigzag edge, the edge-state crossing shifts to $k < \pi$, indicating a negative chirality in Fig.~\ref{edgestate_zigzag}(d).

Although the number of edge-state crossings remains the same in each case, the direction of propagation distinguishes distinct topological phases in parameter space. To identify the edge states more clearly, we choose slightly larger values of $\Gamma_{\scriptscriptstyle \mathcal{K}}$ and $\Gamma_{\scriptscriptstyle \mathcal{K}}^\prime$, which enhance the bulk gap that makes the chiral edge states more prominent. In the present analysis, the phase is tuned by reversing the sign of the DMI vectors. A similar topological transition can also be induced by reversing the direction of the external magnetic field.

\subsection{Magnon orbital moment and orbital Berry curvature}
\label{Orbital_Berry_curvature_zigzag}
\begin{figure*}[t] 
    \centering
    \includegraphics[width=1\linewidth]{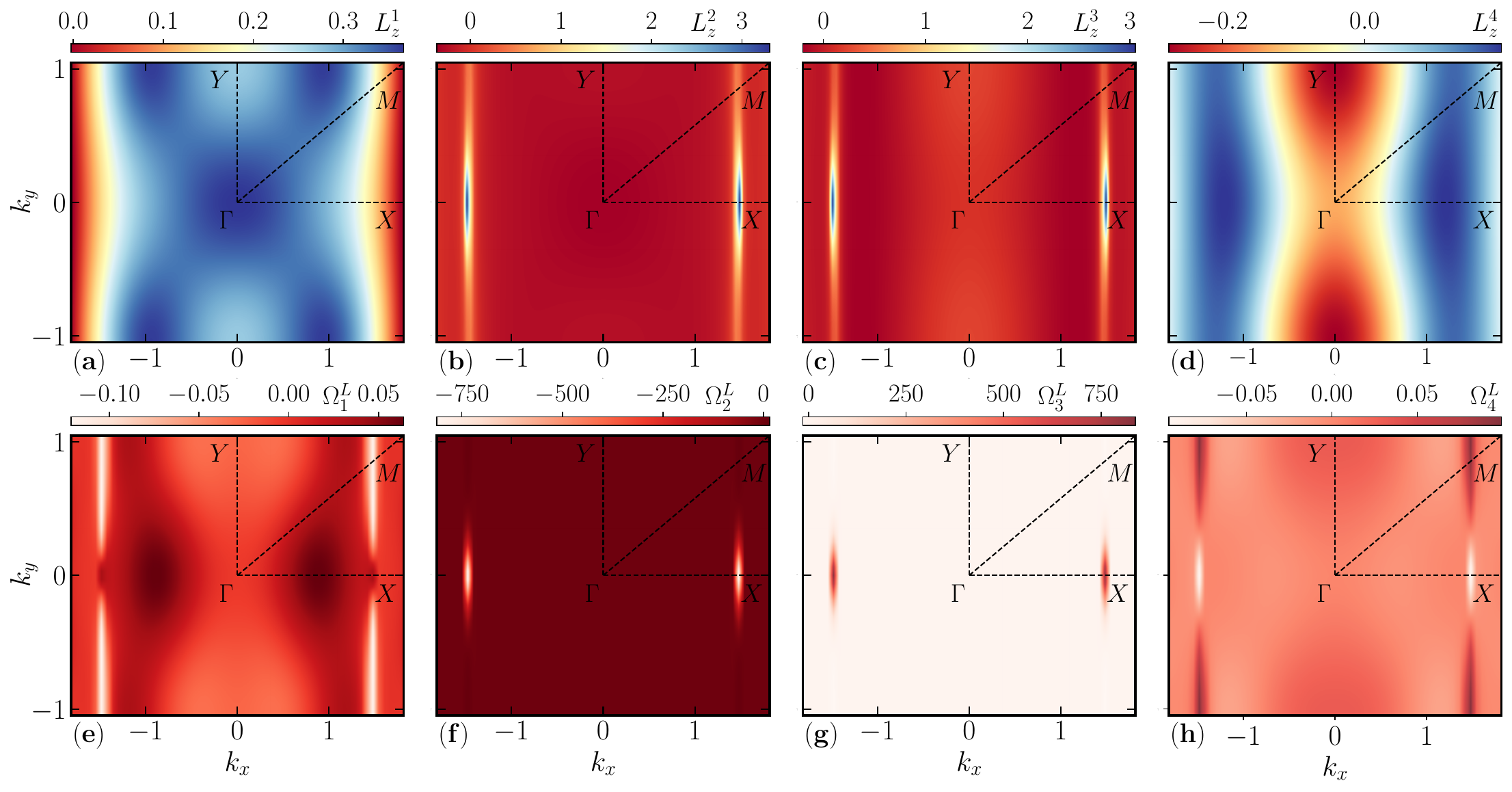}
    \caption{The orbital magnetic moments evaluated over the Brillouin zone for bands (a) $n = 1$, (b) $n = 2$, (c) $n = 3$, and (d) $n = 4$ are shown. Their respective orbital Berry curvatures are illustrated in Figs.~(e)–(h).}
    \label{OBC_zigzag}
\end{figure*}
So far, we have discussed topological signatures such as nonzero Berry curvature and chiral edge states, which have direct implications for the THE. In addition to the THE, another key motivation of our work is to investigate how hybridization between different spin sectors influences another thermal transport phenomenon, namely MONE. Accordingly, in this section we focus on the orbital magnetic moment and the orbital Berry curvature, which play a central role in studying the MONE. 

The total magnetic moment of the system compries of two contributions: one arising from the spin degree of freedom and the other originating from the band structure dependent geometry of magnon wave packets. The latter is referred to as the magnon orbital moment. In a collinear antiferromagnet, the net spin moment, or the magnetization, vanishes, however, the magnon orbital moment may persist. Analogous to the intrinsic orbital moment in electronic systems, one can define the magnon orbital moment operator as~\cite{Go2024},
\begin{equation}
    \hat{L} = \frac{1}{4}(\mb{r} \times \mb{v} - \mb{v} \times \mb{r}),
\end{equation}
whose expectation value is given by,
\begin{equation}
    L_z^n(\mb{k}) = -\frac{i}{2 \hbar}\bra{\partial_k \psi_n}\times\left(\mathcal{H}(\mb{k}) - \eta \bar{\varepsilon}_{n, \mb{k}}\right)\ket{\partial_k \psi_n},
\end{equation}
Here, $n$ denotes the band index, and $\bar{\varepsilon}_{n,\mathbf{k}}$ represents the $n^\text{th}$ eigenvalue of $\eta \mathcal{H}(\mathbf{k})$. In Fig.~\ref{OBC_zigzag}, the magnon orbital moment $L_z^{n}(\mathbf{k})$ is shown for all the four bands. For the lower band ($n = 1$), small yet sharp positive values of the orbital moment are observed near the $\Gamma$ point and between the $M$ and $Y$ symmetry points, while a nearly zero, but weakly negative orbital moment appears along the $M$–$X$ symmetry path as shown in Fig.~\ref{OBC_zigzag}(a).
Similarly, for the upper band, similar to the lower band, which does not directly participate in gap opening induced by hybridization, a very small negative orbital moment is observed near the $Y$ point. As one approaches the Brillouin-zone boundary towards the $M$–$X$ path, an accumulation of the positive values of the orbital moment emerges in Fig.~\ref{OBC_zigzag}(d). 
Moreover, for the middle bands ($n = 2, 3$), the orbital moment accumulation exhibits very similar features for both the bands, differing only in magnitude, as shown in Figs.~\ref{OBC_zigzag}(b) and~\ref{OBC_zigzag}(c).

Further, the Berry curvature, together with the magnon orbital moment, gives rise to another important quantity known as the magnon orbital Berry curvature (MOBC), which can be expressed as~\cite{Go2024},
\begin{equation}
    \Omega_n^L(\mb{k}) = - \sum_{m \neq n} \eta_{mm} \eta_{nn}\frac{2 \hbar~\text{Im} [\bra{n}j_{z,y}^L\ket{m}\bra{m}v_x\ket{n}]}{\left(\bar{\varepsilon}_{n,k} - \bar{\varepsilon}_{m,k}\right)^2}.
\end{equation}
Here, $\hat{j}_{z,y}^L = \left( v_y \eta L_z^{n} + L_z^{n} \eta v_y \right)/4$ denotes the magnon orbital moment current operator, $v_i = -\frac{1}{\hbar}\frac{\partial \mathcal{H}(\mathbf{k})}{\partial k_i}$ ($i = x, y$) represents the velocity operator and $\eta$ is defined earlier. 
Although, the MOBC arises from the combined contribution of the Berry curvature and the magnon orbital moment, all three quantities are odd under time-reversal symmetry. Hence, in a time-reversal-symmetric system, even if the Berry curvature vanishes, the MOBC may still remain finite~\cite{Lee2025}.

In Figs.~\ref{OBC_zigzag}(e–h), we present the MOBC distribution over the MBZ. For the lower band ($n = 1$), shown in Fig.~\ref{OBC_zigzag}(e), a positive MOBC is observed near the band-extrema points associated with the second and third bands, arising from hybridization effects. In contrast, for the upper band ($n = 4$), shown in Fig.~\ref{OBC_zigzag}(h), negative values of MOBC appear near the same regions. Similar to the magnon orbital moment, the MOBC respects the $C_4$ symmetry throughout the MBZ.

Furthermore, Figs.~\ref{OBC_zigzag}(f) and \ref{OBC_zigzag}(g) show that the MOBC acquires large negative and positive contributions at the band-extrema points for the $n = 2$ and $n = 3$ bands, respectively. This behaviour is consistent with our earlier discussion that the MOBC is proportional to the product of the magnon orbital moment, which exhibits similar distributions for both the bands across the MBZ and the Berry curvature, which has opposite signs for the two bands. Consequently, the MOBC acquires opposite signs for the corresponding bands.
However, for completely nondegenerate bands, such a comparison is more straightforward than in cases where degeneracies appear within the MBZ.

In the presence of an external magnetic field, the degeneracy between the up- and down-spin sectors is lifted, and the inclusion of DMI and extended Kitaev interactions leads to hybridization within the MBZ. The resulting small gaps, accompanied by large values of band curvature, are responsible for the enhanced values of the MOBC at the corresponding band extrema. Moreover, since the gap opens primarily between the two middle bands, the magnitudes of both the MOBC and the magnon orbital moment are particularly large for these bands.

\subsection{Transport properties: Thermal Hall effect and magnon orbital Nernst conductivity}
\label{Transport_zigzag}
Thus far we have examined how the presence of an external magnetic field breaks the effective time-reversal symmetry of a zigzag-ordered antiferromagnet, and how the inclusion of DMI and extended Kitaev interactions induces hybridization between different spin sectors via spin-excitation quasiparticles, namely magnons. In the presence of these interactions, we have characterized the topology of the bands using mathematically well-defined quantities that may not directly be accessible to experimental validation.
Accordingly, in this section we focus on experimentally accessible observables that can be directly compared with measurements, thereby offering robust support of our theoretical framework and underscoring the completeness of the present study.
Therefore, we identify two types of transverse transport phenomena, namely, the thermal Hall conductivity (THC) and the magnon orbital Nernst conductivity (MONC). These are associated with the thermal current ($j^{Q}$) and the orbital moment current ($j^{\mathrm{OM}}$), respectively, and are defined as
\begin{subequations}
\begin{align}
j_y^{Q} &= -\kappa_{xy}\partial_x T, \\
j_y^{\mathrm{OM}} &= -\alpha_{xy}\partial_x T .
\end{align}
\end{subequations}
Here, $y$ denotes the direction of the induced transverse current in response to a temperature gradient $\partial_x T$ applied along the $x$-direction. The proportionality coefficients $\kappa_{xy}$ and $\alpha_{xy}$ correspond to the THC and MONC, respectively, and are given by~\cite{Debnath2025, Go2024},
\begingroup
\allowdisplaybreaks
\begin{subequations}
    \begin{align}
        \kappa_{xy} &= - \frac{k_B T}{4 \pi^2}\sum_n\int_{\tiny{BZ}} c_2(\rho_{n, \mb{k}}) \Omega_n^z (\mb{k}) \cdot d\mb{k},\label{THC_zigzag}\\
        \alpha_{xy} & = \frac{2k_B}{\hbar V}\sum_n\int_{\tiny{BZ}} c_1 (\rho_{n, \mb{k}}) \Omega_n^L(\mb{k}). d \mb{k}\label{MONC_zigzag},
    \end{align}
\end{subequations}
\endgroup
where $k_B$ is the Boltzmann constant, $T$ is the absolute temperature and $\rho_{n, \mb{k}} = 1/[\exp{\varepsilon_n(\mb{k})/k_B T} - 1]$ is the Bose Einstein distribution function and $c_i$s ($i = 1,2$) are the weighting functions defined by
\begin{subequations}
    \begin{align}
        c_1(\rho_{n, \mb{k}}) & = (1 + \rho_{n, \mb{k}})\ln{(1 + \rho_{n, \mb{k}})}- \rho_{n, \mb{k}}\ln{\rho_{n, \mb{k}}},\\
        c_2(\rho_{n, \mb{k}}) & = (1 + \rho_{n,\mathbf{k}})\left(\log{\frac{1 + \rho_{n,\mathbf{k}}}{\rho_{n,\mathbf{k}}}}\right)^2
    - (\log{\rho_{n,\mathbf{k}}})^2 \nonumber\\
    &- 2\text{Li}_2 (- \rho_{n,\mathbf{k}}).
    \end{align}
\end{subequations}
Here, $\mathrm{Li}_2(-\rho_{n,\mathbf{k}})$ represents the polylogarithmic function. At low temperatures, the weighting functions $c_1(\rho_{n,\mathbf{k}})$ and $c_2(\rho_{n,\mathbf{k}})$ are strongly enhanced at low energies. As a result, at low temperatures, both the THC and MONC are predominantly governed by low-energy magnon bands, and the contributions from higher-energy bands are effectively suppressed~\cite{An2025,Debnath2025}.

\begin{figure*}[t]
    \centering
    \includegraphics[width=0.8\linewidth]{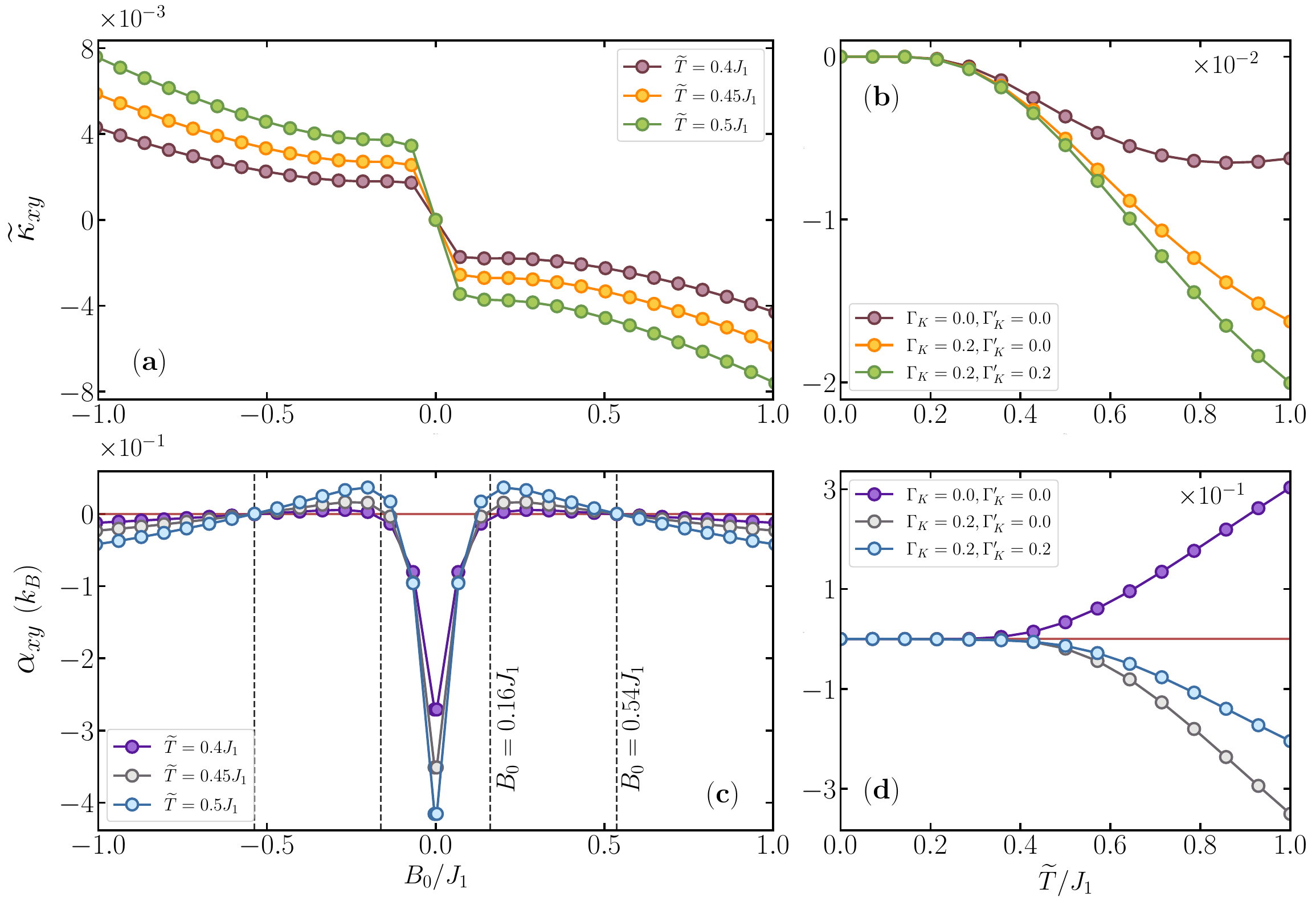}
    \caption{In a zigzag antiferromagnet, the thermal Hall conductivity is shown as a function of (a) the external magnetic field $B_0$ and (b) the temperature $\tilde{T}$. The magnon orbital Nernst conductivity as a function of (c) $B_0$ and (d) $\tilde{T}$ are also presented. Here, $\tilde{T} = k_B T$. In panels (a) and (c), the parameters $\Gamma_{\scriptscriptstyle \mathcal{K}} = \Gamma_{\scriptscriptstyle \mathcal{K}}^\prime = 0.2~\text{meV}$ are used, while in panels (b) and (d) the external magnetic field is set to $B_0 = 1~\text{meV}$. All other parameters are fixed at $J_1 = 1.49~\text{meV}$, $J_2 = 0.04~\text{meV}$, $J_3 = -0.6~\text{meV}$, $\Delta = -3.6~\text{meV}$, $D = 0.3~\text{meV}$, and $\mathcal{K} = 0.4~\text{meV}$.}
    \label{THC_MONC_zigzag}
\end{figure*}
In this study, we examine the behaviour of $\kappa_{xy}$ and $\alpha_{xy}$ as functions of two external parameters, namely temperature ($T$) and the external magnetic field ($B_0$). All other parameters, including $J_m$, $D$, $\mathcal{K}$, and related couplings, are intrinsic to a particular material. Therefore, we choose a set of parameters $J_1 = 1.49~\text{meV}$, $J_2 = 0.04~\text{meV}$, $J_3 = -0.6~\text{meV}$, $\Delta = -3.6~\text{meV}$, $\mathcal{K} = 0.4~\text{meV}$, $D = 0.3~\text{meV}$, and $\Gamma_{\scriptscriptstyle \mathcal{K}} = \Gamma_{\scriptscriptstyle \mathcal{K}}^\prime = 0.2~\text{meV}$ such that the zigzag ordering in the collinear antiferromagnetic system remains stable as the ground state. In addition, the presence of DMI and Kitaev interactions ensures hybridization between different spin sectors, allowing the behaviour of these quantities to be studied within the topological regime. In keeping with ongoing spirit of the discussion, we first discuss THE which has a topological origin, while the MONE, which does not directly depend upon topology, is taken up next.

In Fig.~\ref{THC_MONC_zigzag}(a), we present the thermal Hall conductivity $\kappa_{xy}$ as a function of the external magnetic field $B_0$. For positive values of $B_0$, and for the chosen parameter set, the total Chern number associated with the lower two bands is $+1$. Although the THC formally includes contributions from all the bands, at low temperatures the dominant contribution arises from the lowest band. Consequently, the integral in Eq.~\eqref{THC_zigzag} is governed by the sign of the Chern number of the lower bands. Thus, for $k_B T/J_1 = 0.4$, $0.45$, and $0.5$, we observe negative values for $\kappa_{xy}$, consistent with the overall sign appearing in Eq.~\eqref{THC_zigzag}.

Further, in the absence of an external magnetic field ($B_0 = 0$), the system preserves effective TRS. Since the time-reversal operator reverses the thermal Hall current as $j_y^{Q} \rightarrow -j_y^{Q}$, the net thermal Hall current, and consequently the THC must vanish. This behaviour is clearly observed in Fig.~\ref{THC_MONC_zigzag}(a).
Although local Berry curvature may remain finite at certain points within the MBZ when $B_0 = 0$, its overall contribution integrates to zero. This highlights that, despite the enhancement of the THC due to band hybridization and nonzero Chern numbers, a finite Hall response cannot be realized without the breaking of time-reversal symmetry, that may occur either through an external magnetic field in a collinear antiferromagnet or via a net magnetization in a canted antiferromagnet (or a ferromagnet). Moreover, reversing the direction of $B_0$ inverts the direction of the Hall current, leading to a corresponding sign change in $\kappa_{xy}$ for negative values of $B_0$.

In Fig.~\ref{THC_MONC_zigzag}(a), we also observe that the magnitude of $\kappa_{xy}$ increases with temperature, although this increase is not linear. To further elucidate this behaviour, Fig.~\ref{THC_MONC_zigzag}(b) shows the temperature dependence of $\kappa_{xy}$. For $\mathcal{K} = 0.4~\text{meV}$ and $D = 0.3~\text{meV}$, the system resides in a topological phase characterized by a Chern number of $-1$ for the two lower bands.
Although the energy separation between the extrema of the two middle bands is very small, leading to large Berry curvature contributions due to band hybridization, the corresponding weighting function $c_2(\rho_{n,\mathbf{k}})$ remains nearly identical for these bands. Consequently, the dominant contribution to $\kappa_{xy}$ originates from the lowest-energy band, for which the Berry curvature is predominantly positive across the MBZ. As a result, $\kappa_{xy}$ assumes negative values, consistent with the sign convention in Eq.~\eqref{THC_zigzag}.
As the temperature increases, $\kappa_{xy}$ deviates from the linear behaviour. This nonlinearity arises because the weighting function $c_2(\rho_{n,\mathbf{k}})$ acquires enhanced contributions from the high-energy bands, which host Berry curvatures of opposite signs. These competing contributions lead to the observed temperature dependence of $\kappa_{xy}$, as shown in Fig.~\ref{THC_MONC_zigzag}(b).

In contrast, the presence of off-diagonal interactions arising from the extended Kitaev terms, specifically for $\Gamma_{\scriptscriptstyle \mathcal{K}} = 0.2~\text{meV}$, $\Gamma_{\scriptscriptstyle \mathcal{K}}^\prime = 0$, and for $\Gamma_{\scriptscriptstyle \mathcal{K}} = \Gamma_{\scriptscriptstyle \mathcal{K}}^\prime = 0.2~\text{meV}$, the Berry curvature associated with the two lower bands renders to the predominantly positive, yielding a combined Chern number equal to $+1$. In addition, the energy gap between the two middle bands at the band extrema points along the symmetry paths increases. Consequently, the temperature dependence of $\kappa_{xy}$ becomes nearly linear and remains negative.

Owing to the different underlying mechanism, the magnon orbital current exhibits features that are distinct from those of the thermal Hall current, and so does the magnon orbital Nernst conductivity $\alpha_{xy}$. In Fig.~\ref{THC_MONC_zigzag}(c), we show the variation of $\alpha_{xy}$ as a function of the external magnetic field $B_0$. Unlike the THC ($\kappa_{xy}$), $\alpha_{xy}$ does not vanish at $B_0 = 0$, instead, it attains relatively large values in this case.
In contrast to the THE, the magnon orbital Nernst effect is an intrinsic property of the system. The orbital Nernst current remains finite even in the presence of time-reversal symmetry, as it carries a net transverse orbital moment arising from contributions of different spin sectors. Consequently, a finite $\alpha_{xy}$ is observed at $B_0 = 0$.
Furthermore, at $B_0 = 0.16J_1$, we observe a sign change in $\alpha_{xy}$. Similar to the weighting function $c_2(\rho_{n,\mathbf{k}})$, $c_1(\rho_{n,\mathbf{k}})$ also exhibits a dominant contribution arising primarily from the lowest energy band, thereby directly determining its sign. However, unlike the Berry curvature, the orbital Berry curvature does not retain an integer-value upon integration over the Brillouin zone. For larger values of $B_0$, namely, at $B_0 = 0.54J_1$, $\alpha_{xy}$ again becomes negative and subsequently increases approximately linearly with increasing $B_0$.
In the absence of an external magnetic field, the different spin sectors cannot be separated, and spin mixing induced by DMI or Kitaev interactions enhances the MONC. However, upon applying an external magnetic field, the energy separation between the spin sectors increases, which in turn suppresses the orbital Nernst response.

With regard to the thermal effect, similar to the THC, the MONC also increases with temperature. In Fig.~\ref{THC_MONC_zigzag}(d), we present the temperature dependence of $\alpha_{xy}$. In the absence of off-diagonal interactions, $\alpha_{xy}$ remains positive and increases monotonically with increasing temperature.
Upon introducing $\Gamma_{\scriptscriptstyle \mathcal{K}}$, $\Gamma_{\scriptscriptstyle \mathcal{K}}^\prime$, or both, the system undergoes a change in its topological state, as discussed earlier. Unlike the THC, these distinct topological phases are more discernly manifested in the behaviour of the MONC, as shown in Fig.~\ref{THC_MONC_zigzag}(d). Additionally, for $\Gamma_{\scriptscriptstyle \mathcal{K}} = 0.2~\text{meV}$ and $\Gamma_{\scriptscriptstyle \mathcal{K}}^\prime = 0$, the magnitude of $\alpha_{xy}$ is larger than in the case where $\Gamma_{\scriptscriptstyle \mathcal{K}} = \Gamma_{\scriptscriptstyle \mathcal{K}}^\prime = 0.2~\text{meV}$.
A possible origin of this enhancement is that these bond-dependent interactions modify the band geometry along the high-symmetry paths, leading to an evolution of the MOBC that becomes strongly enhanced in the vicinity of the low-energy sectors. Since low-energy magnons dominate the transport response, we observe in comparatively larger values of $\alpha_{xy}$.

\begin{figure}[b]
    \centering
    \includegraphics[width=1\linewidth]{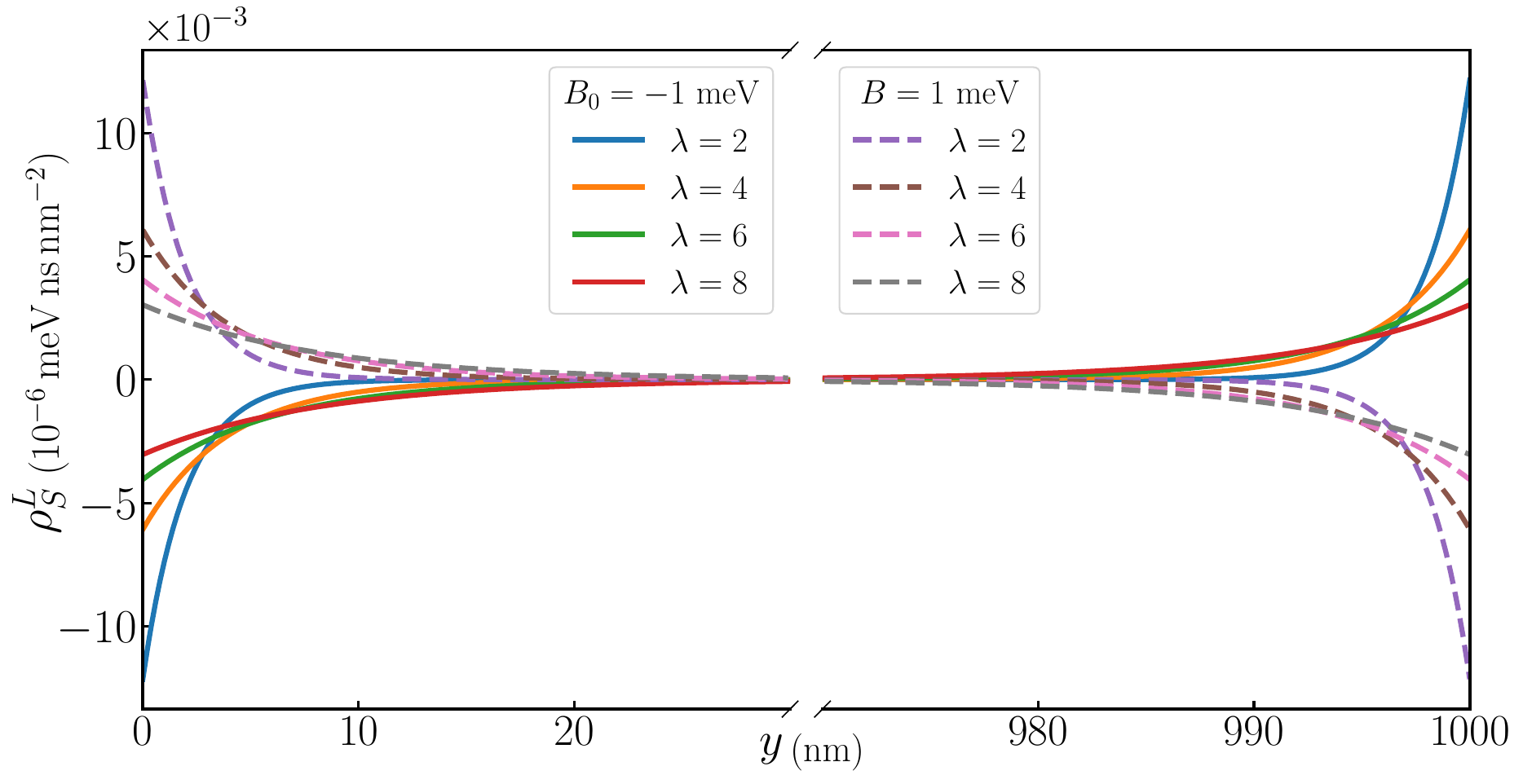}
    \caption{Spin-polarized magnon orbital moment accumulations along one of the spatial directions, namely the $y$-direction under an applied magnetic field $B_0 = \pm 1~\mathrm{meV}$ have been shown.}
    \label{polarization_zigzag}
\end{figure}
In addition to the total MONC, one can also define a spin-polarized magnon orbital Nernst conductivity, given by, $\alpha_{xy}^s = \alpha_{xy}^\alpha - \alpha_{xy}^\beta$, where $\alpha \in {A_1, A_4}$ and $\beta \in {B_2, B_4}$. In the absence of an external magnetic field ($B_0 = 0$), this quantity vanishes, since the up- and down-spin sectors remain unresolved.
When the DMI and Kitaev interactions are present, the hybridization between different spin sectors opens a gap in the magnon spectrum. In this regime, the orbital Berry curvature becomes large due to the mixing of different spin states, which makes it difficult to clearly separate the MONC contributions from individual spin sectors. However, at sufficiently low temperatures, the dominant contribution arises from the lowest-energy band, which corresponds predominantly to a single spin sector, determined by the direction of the applied magnetic field.
Consequently, in the low-temperature limit, one can obtain an approximate representation of the spin-polarized magnon orbital moment accumulation, defined as $\rho_s^L = \rho_\alpha^L - \rho_\beta^L$, which can be expressed via~\cite{Go2024},
\begin{equation}
    \rho_s^L(y) = \alpha_{xy}^s \frac{\tau \sinh{\left(\frac{W - 2y}{2\lambda}\right)}}{\lambda \cosh{\left(\frac{W}{2\lambda}\right)}} \partial_x T.
\end{equation}
Here $\tau$ denotes the magnon orbital relaxation time, $\lambda$ is the diffusion length, and $W$ represents the width of the sample. For $k_B T/J_1 = 0.2$, and $B_0 = \pm 1~\text{meV}$ $\alpha_{xy}^s \sim \mp 3.26\times 10^{-4} k_B$, the corresponding spin-polarized magnon orbital moment accumulation is shown in Fig.~\ref{polarization_zigzag}. Here, $\rho_s^L$ changes sign upon reversing the direction of $B_0$, leading to accumulation of orbital moments with opposite signs at the two boundaries. This behaviour is analogous to electronic polarization in conventional systems~\cite{Go2024}.

\subsection{Qualitative overview of the Néel antiferromagnetic state}
\label{Neel_antiferromagent}
To this end, we have presented a detailed analysis of the topological characterization of a zigzag antiferromagnet and the transport features, such as, THC and MONC. Moreover, in addition to the antisymmetric and bond-dependent interactions, the zigzag ordering in a collinear antiferromagnet itself plays an important role in realizing topological states through hybridization between different spin sectors. Therefore, for the sake of completeness, it is instructive to compare those results to a N\`eel-ordered honeycomb antiferromagnet. The spin Hamiltonian for this system follows the same form as that given in Eq.~\eqref{Model_Hamiltonain_zigzag}, but the Heisenberg exchange interactions are restricted to be nearest neighbours, which can be achieved by setting $J_2 = J_3 = 0$. In addition, we choose positive values for the anisotropy term $\Delta$, that ensures the N\`eel-ordered state to remain energetically stable.

\begin{figure}[t]
    \centering
    \includegraphics[width=1\linewidth]{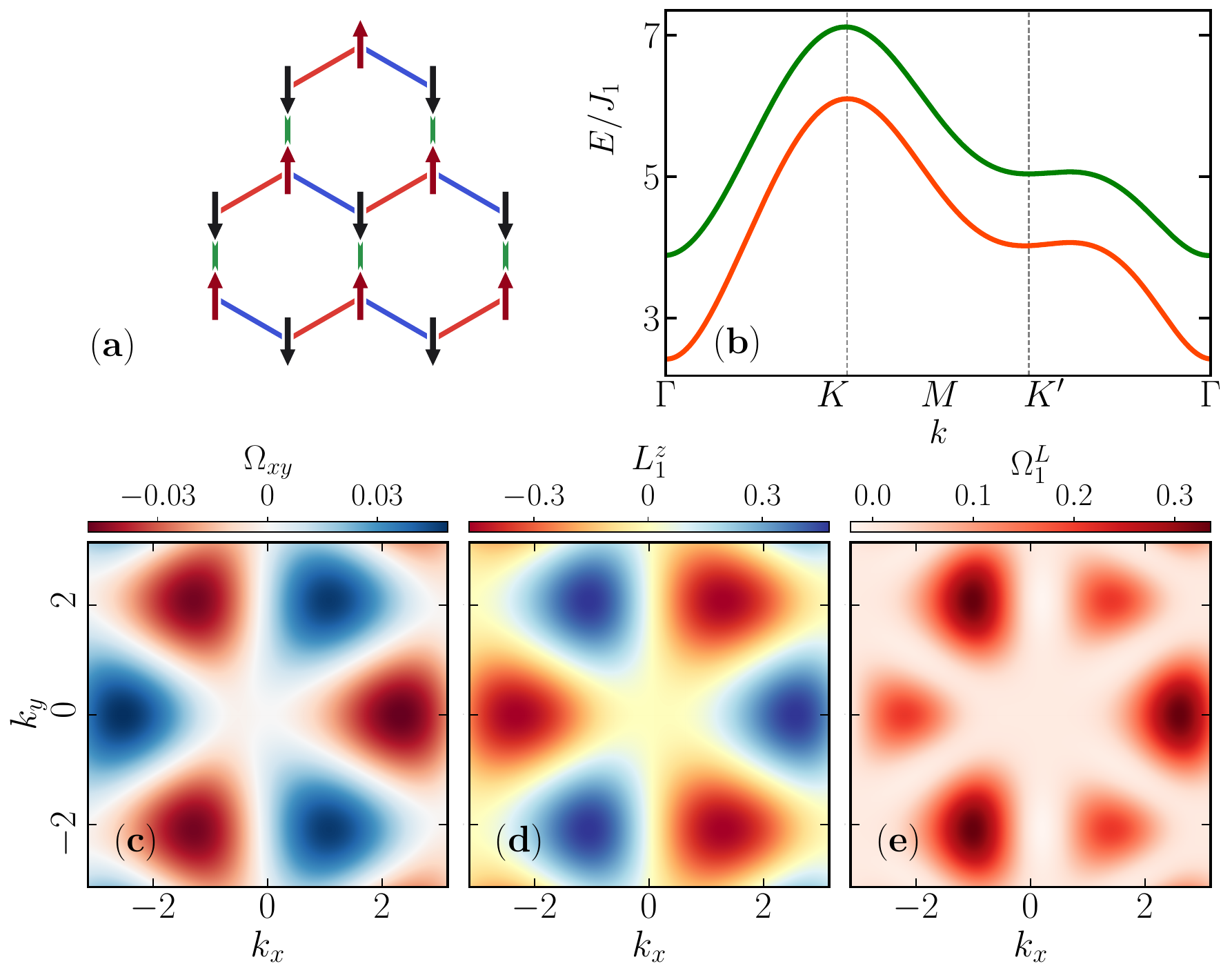}
    \caption{(a) Schematic diagram of the N\'eel antiferromagnet has been shown, where the red and black arrows represent up and down spins, respectively. (b) Bulk band structure has been shown in presence of external magnetic field, with $B_0 = 0.5~\text{meV}$, $\Delta= 0.5~\text{meV}$, $J_1 = 1.49~\text{meV}$, $D = 0.2~\text{meV}$, $\mathcal{K} = \Gamma_{\scriptscriptstyle \mathcal{K}} = \Gamma_{\scriptscriptstyle \mathcal{K}}^\prime = 0.1~\text{meV}$. Accordingly, (c) the Berry curvature, (d) magnon orbital moment, and (e) the orbital Berry curvature have been shown corresponding to the lower band.}
    \label{BC_OBC_Honeycomb}
\end{figure}

In this case, the system consists of two sublattices, namely, sublattice $A$($B$) with up(down) spins, as shown in Fig.~\ref{BC_OBC_Honeycomb}(a). After performing the spin–boson transformation as described in Eq.~\eqref{HP_zigzag}, the magnon Hamiltonian is written in Appendix~\ref{Neel_Hamiltonian}, and the corresponding magnon band structure is shown for $J_1 = 1.49~\text{meV}$, $\mathcal{K} = 0.1~\text{meV}$, $D = 0.2~\text{meV}$, and $\Gamma_{\scriptscriptstyle \mathcal{K}} = \Gamma_{\scriptscriptstyle \mathcal{K}}^\prime = 0.1~\text{meV}$. The Brillouin zone associated with the Néel-ordered configuration is hexagonal, with the high-symmetry points being $\Gamma$, $K$, $K'$, and $M$. Unlike the zigzag-ordered case, no band inversion is observed along the high-symmetry paths as shown in Fig.~\ref{BC_OBC_Honeycomb}(b).

Furthermore, we present the Berry curvature corresponding to the lower band, which exhibits a $C_3$ symmetry arising from the valley asymmetry at the $K$ and $K'$ points. Although the application of an external magnetic field $B_0$ breaks the time-reversal symmetry and generates a finite Berry curvature within the Brillouin zone, however, the integrated contribution remains zero, yielding a trivial Chern number $C_n = 0$ for both the bands ($n = 1, 2$).
A similar spatial distribution is observed for the magnon orbital moment, as shown in Fig.~\ref{BC_OBC_Honeycomb}(d). As discussed earlier, the MOBC is proportional to the product of the magnon orbital moment and the Berry curvature. Consequently, the MOBC acquires predominantly positive values throughout the Brillouin zone, as illustrated in Fig.~\ref{BC_OBC_Honeycomb}(e) for the lower band ($n = 1$).
For the upper band ($n = 2$), both the Berry curvature and the magnon orbital moment reverse their signs, however, their product remains positive, resulting in a positive MOBC across the Brillouin zone as well. The observed valley asymmetry originates from the presence of DMI, while the inclusion of the Kitaev interactions shifts the band extrema away from the high-symmetry points $K$ and $K'$.
\begin{figure}[t]
    \centering
    \includegraphics[width=1\linewidth]{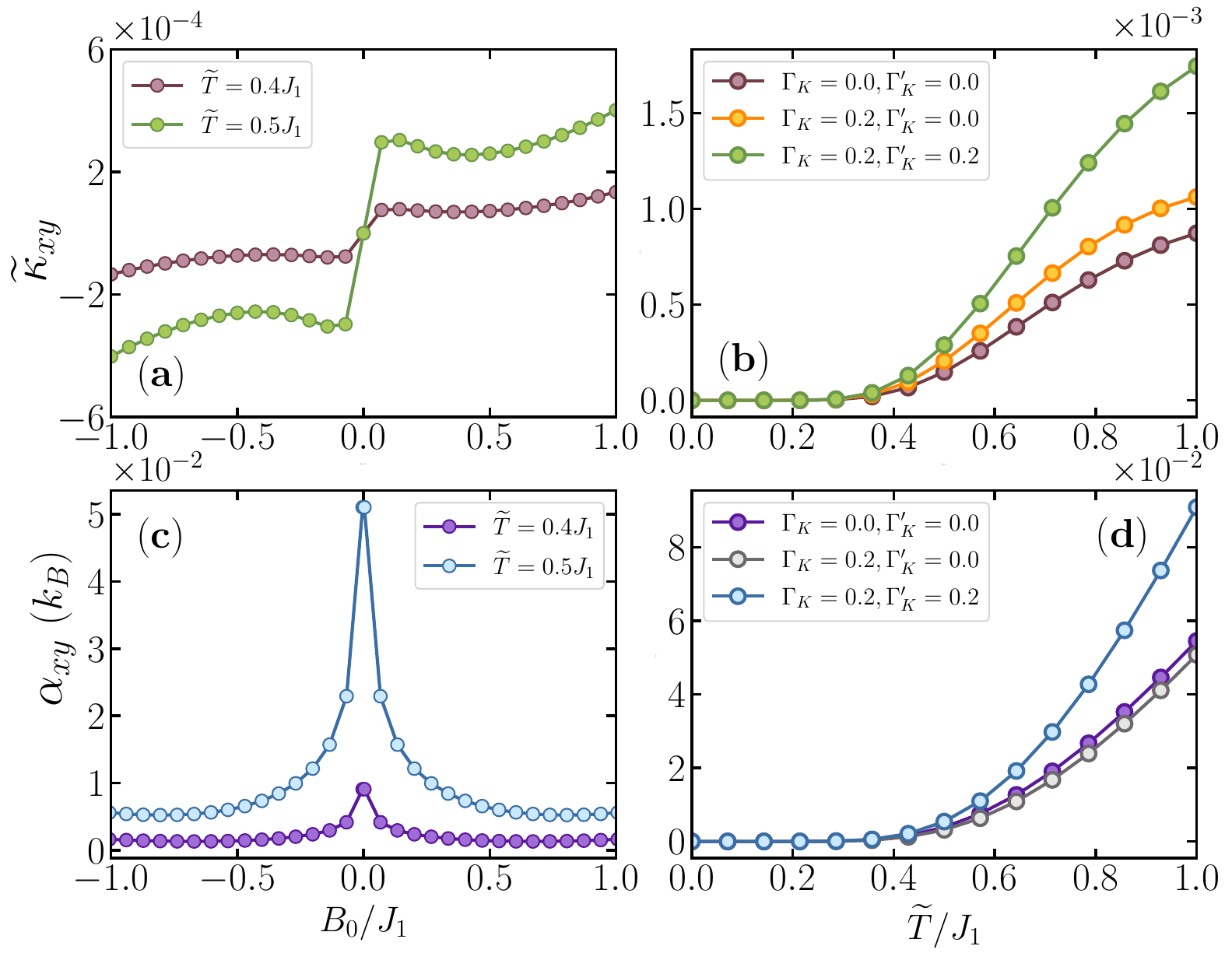}
    \caption{The thermal Hall conductivity for N\'eel antiferromagnet is shown as a function of (a) the external magnetic field $B_0$ and (b) the temperature $\tilde{T}$. Also, The magnon orbital Nernst conductivity is presented as a function of (c) $B_0$ and (d) $\tilde{T}$. Here, $\tilde{T} = k_B T$. In panels (a) and (c), the parameters $\Gamma_{\scriptscriptstyle \mathcal{K}} = \Gamma_{\scriptscriptstyle \mathcal{K}}^\prime = 0.2~\text{meV}$ are used, while in panels (b) and (d) the external magnetic field is set to $B_0 = 1~\text{meV}$. All other parameters are kept fixed at $J_1 = 1.49~\text{meV}$, $\Delta = 1~\text{meV}$, $D = 0.3~\text{meV}$, and $\mathcal{K} = 0.4~\text{meV}$.}
    \label{THC_MONC_Honeycomb}
\end{figure}

We further examine the THC and MONC as functions of the external magnetic field $B_0$ and temperature $T$, using the same parameter values for $\mathcal{K}$ and $D$ as those employed for Fig.~\ref{THC_MONC_zigzag}. In Fig.~\ref{THC_MONC_Honeycomb}(a), we present $\kappa_{xy}$ as a function of $B_0$ at two different temperatures, namely, $\tilde{T} = 0.4$ and $0.5$ in unit of $J_1$. As $B_0$ increases, $\kappa_{xy}$ initially decreases and subsequently increases beyond a certain field strength. This nonmonotonic behaviour becomes more pronounced at higher temperatures. Upon reversing the direction of $B_0$, the sign of $\kappa_{xy}$ changes, as expected.
In addition, we present $\kappa_{xy}$ as a function of temperature. In all cases, $\kappa_{xy}$ increases with increasing temperature, and similar to the zigzag-ordered system shown in Fig.~\ref{THC_MONC_zigzag}(b), its magnitude is enhanced for nonzero values of $\Gamma_{\scriptscriptstyle \mathcal{K}}$ and $\Gamma_{\scriptscriptstyle \mathcal{K}}^\prime$. However, unlike the zigzag-ordered phase, the Néel-ordered state remains topologically trivial with $C = 0$. Consequently, the magnitude of $\kappa_{xy}$ is significantly suppressed as can be seen in Figs.~\ref{THC_MONC_Honeycomb}(a) and~\ref{THC_MONC_Honeycomb}(b).

The MONC as a function of $B_0$ exhibits qualitatively similar behaviour as that of a zigzag antiferromagnet. In general, and in the absence of DMI or Kitaev interactions, the MONC is nearly independent of the external magnetic field, as reported in the literature~\cite{Go2024}. In the present case, however, the inclusion of these interactions introduces a competition with the Zeeman term, leading to a field-dependent MONC. This behaviour further indicates that a significant contribution to the MONC originates from the hybridization between different spin sectors, which is gradually suppressed by band splitting in the presence of an external magnetic field (see Fig.~\ref{THC_MONC_Honeycomb}(c)).
Unlike the zigzag-ordered system, this hybridization does not induce finite Chern number in the Néel-ordered phase. Consequently, as shown in Fig.~\ref{THC_MONC_Honeycomb}(d), $\alpha_{xy}$ increases with temperature without exhibiting features associated with the topological transitions. Moreover, when both $\Gamma_{\scriptscriptstyle \mathcal{K}}$ and $\Gamma_{\scriptscriptstyle \mathcal{K}}^\prime$ are present, the magnitude of $\alpha_{xy}$ is larger than in the case where $\Gamma_{\scriptscriptstyle \mathcal{K}}^\prime = 0$.

\section{Conclusion}
\label{conclusion_zigzag}
We have investigated the topological properties of magnon excitations in a zigzag-ordered collinear antiferromagnet in the presence of various magnetic interactions, including long-range exchange couplings, DMI, and extended Kitaev interactions. While the effects of DMI and an external magnetic field in such systems have been explored previously, the inclusion of extended Kitaev interactions reveals several unique features.
A key motivation of this work is to explore the magnon orbital Nernst effect, which has emerged as an important mechanism governing orbital and electronic polarization phenomena. We show that the nontrivial topological states originate from the hybridization between different spin sectors induced by the combined effects of antisymmetric  and bond dependent interactions in presence of an external magnetic field. In an experimental realizable parameter regime, a detailed analysis of orbital current profiles provides useful insights and uncovers new transport signatures associated with the magnon band topology.

Let us enumerate the main highlights of our work. In the zigzag-ordered ground state of a collinear antiferromagnet, we find two doubly degenerate magnon bands originating from opposite glide-mirror eigenvalues in the absence of an external magnetic field. When a magnetic field is applied, this twofold degeneracy, associated with up- and down-spin sectors is lifted. However, the presence of DMI alone does not cause opening of a band gap.
Upon introducing the Kitaev interaction, band gaps open at several points in the Brillouin zone, rendering topologically nontrivial states characterized by nonzero Chern numbers. We also observe multiple topological phase transitions that occur through gap-closing events as the magnitude of the interactions are tuned within the experimental limits. Furthermore, due to symmetry protected degeneracies along the $M$–$X$ symmetry path, the combined Chern number of the bands is restricted to values of $\pm 1$, despite the presence of multiple bands.
In addition, we demonstrate the emergence of chiral edge states within the topological band gaps. Proceeding beyond a mere topological characterization of the above scenario, we investigate transport features, such as the magnon orbital Nernst conductivity, that is MONC. Notably, unlike the THE, breaking TRS is not a necessary condition for the orbital Nernst response. 
Furthermore, even without an external magnetic field, the combined effects of interband transitions and hybridization between different spin sectors give rise to a giant orbital Nernst conductivity. This response is strongly impeded upon applying an external magnetic field due to the spin splitting, which suppresses interband transitions. However, beyond a critical field strength, the Nernst conductivity begins to increase slowly again, as hybridization between spin sectors supersedes over the interband-transition effects. Although the THC is unable to clearly distinguish between the different phases in this case, the MONC provides a clear signature of the distinct topological phases. 
In addition, for a representative set of parameter values (that also have experimental relavance), we explicitly present two key quantities responsible for the Nernst conductivity, namely, the magnon orbital moment and the MOBC corresponding to all the four magnon bands for a zigzag antiferromagnet. The orbital moment, originating from the magnon wave packet is known to contribute to various orbitronic properties. We have demonstrated the accumulation of the orbital moment along the system edges, which gives rise to the corresponding Nernst current associated with the Nernst conductivity. This phenomenon is analogous to the electronic polarization~\cite{Go2024,To2024}, suggesting that magnonic properties can also be influenced and manipulated via an electric field. Altogether, these theoretical findings will enrich studies on the potential of magnons in orbitronics and modern quantum device applications.

For completeness and to highlight the distinct role plaed by the zigzag ordering, we have also briefly discussed the Néel-ordered state. A key difference between the two cases is that, in the Néel-ordered configuration, the combined effects of the Kitaev interaction and DMI are insufficient to induce a topological gap and hence renders zero Chern numbers. In contrast, the zigzag-ordered state supports the opening of nontrivial band gaps and the emergence of topological phases, which in turn are responsible for the enhanced thermal Hall conductivity and the larger orbital Nernst response.
\appendix 
\begin{widetext}
\section{Real-Space Spin Hamiltonian of Zigzag Ordered Antiferromagnet in the Magnon Basis}
\label{spin_Hamiltonian_zigzag}
A spin-boson transformation following Eq.~\eqref{HP_zigzag}, the Hamiltonian in the magnon basis can be written as,
    \begin{equation}
        \begin{split}
            H &= (-J_1 + 3J_3 + K + \Delta + \mu_B B_0)\sum_{i}\left[ a_{1i}^\dagger a_{1j} + a_{4i} ^\dagger a_{4j}\right]
            + (-J_1 + 3J_3 + K + \Delta - \mu_B B_0)\sum_{i} \left[b_{2i}^\dagger b_{2j} + b_{3i}^\dagger b_{3j}\right]\\ 
            &+J_1S\sum_{\braket{i,j}}\left[a_{1i}^\dagger b_{2j} + a_{1i}^\dagger b_{2j}^\dagger + a_{4i}^\dagger b_{3j} + a_{4i}^\dagger b_{3j}^\dagger + b_{2i}^\dagger b_{3j} + a_{1i}^\dagger a_{4j}+ h.c.\right]+ J_2S\sum_{\braket{\braket{i,j}}}\left[a_{1i}^\dagger b_{3j} +  a_{1i}^\dagger b_{3j}^\dagger + a_{4i}^\dagger b_{2j}\right.\\
            &+\left.  a_{4i}^\dagger b_{2j}^\dagger + a_{1i}^\dagger a_{1j} + b_{2i}^\dagger b_{2j} + b_{3i}^\dagger b_{3j} + a_{4i}^\dagger a_{4j} + h.c.\right] + J_3 S \sum_{\braket{\braket{\braket{i,j}}}} \left[a_{1i}^\dagger b_{2j} + a_{1i}^\dagger b_{2j}^\dagger + a_{4i}^\dagger b_{3j} + a_{4i}^\dagger b_{3j}^\dagger + h.c.\right] \\
            &+ D \sum_{\braket{\braket{i,j}}}i \nu_{ij}\left[a_{1i}b_{3j} + b_{3i}^\dagger a_{1j}^\dagger + a_{4i}b_{2j} + b_{2i}^\dagger a_{4j}^\dagger + a_{1i}^\dagger a_{1j} + b_{2i}^\dagger b_{2j} + b_{3i}^\dagger b_{3j} + a_{4i}^\dagger a_{4j} + h.c. \right]\\
            & + \sum_{\braket{i,j}^x} \left[\frac{\mathcal{K}S}{2}\left(b_{2i}^\dagger b_{3j} + b_{2i}^\dagger b_{3j}^\dagger + a_{1i}^\dagger a_{4j} + a_{1i}^\dagger a_{4j}^\dagger\right) + \i \Gamma_{\scriptscriptstyle \mathcal{K}}^{\prime}S\left(a_{1i}^\dagger a_{4j}^\dagger - b){2i}^\dagger b_{3j}^\dagger\right) + h.c.\right] + \sum_{\braket{i,j}^y} \left[\frac{\mathcal{K}S}{2}\left(b_{2i}^\dagger b_{3j} \right.\right.\\
            &\left.\left.- b_{2i}^\dagger b_{3j}^\dagger + a_{1i}^\dagger a_{4j} - a_{1i}^\dagger a_{4j}^\dagger\right)+ i \Gamma_{\scriptscriptstyle \mathcal{K}}^{\prime}S\left(a_{1i}^\dagger a_{4j}^\dagger - b){2i}^\dagger b_{3j}^\dagger\right) + h.c.\right] + \sum_{\braket{i,j}^z}\left[i\Gamma_{\scriptscriptstyle \mathcal{K}} S \left(a_{1i}^\dagger b_{2j} + a_{4i}^\dagger b_{3j}\right) + h.c.\right],
        \end{split}\label{real_H_zigzag}
    \end{equation}
where $a_1,~b_2,~b_3,~a_4$ are the bosonic operators corresponding to the sublattices $A_1,~B_2,~B_3,~\text{and}~A_4$. Also in the last three terms, $\braket{i,j}^\gamma$ with $\gamma= x, y, z$ correspond to the $x,~y$ and $z$-bonds as depicted in Fig.~\ref{Model_zigzag}. 
\section{Model Hamiltonian For N\'eel Ordered Antiferromagnet}
\begingroup
\allowdisplaybreaks
\label{Neel_Hamiltonian}
The spin Hamiltonian corresponding to the Néel-ordered state can be written in the same form as Eq.~\eqref{Model_Hamiltonain_zigzag}, with $n = 1$, indicating that only nearest-neighbour exchange interactions are included. Following Eq.~\eqref{HP_zigzag}, we perform a spin–boson transformation for the $A$ and $B$ sublattices, which yields the transformed BdG Hamiltonian in the basis $(a_{\mathbf{k}},, b_{\mathbf{k}},, a_{-\mathbf{k}}^\dagger,, b_{-\mathbf{k}}^\dagger)^{T}$, given by,
\begin{equation}
    \mathcal{H}_N(\mb{k}) = \begin{bmatrix}
                                \bar{M}_A(\mb{k}) & \bar{f}_1(\mb{k}) & 0 & \bar{f}_2(\mb{k})\\
                                \bar{f}^*_1(\mb{k}) & \bar{M}_B(\mb{k}) & \bar{f}_2(-\mb{k}) & 0\\
                                0 & \bar{f}^*_2(-\mb{k}) & \bar{M}_A(-\mb{k}) & \bar{f}^*_1(-\mb{k})\\
                                \bar{f}^*_2(\mb{k}) & 0 & \bar{f}_1(-\mb{k}) & \bar{M}_B(-\mb{k})
                            \end{bmatrix}
\end{equation}
where 
\begin{subequations}
\begin{align}
    \begin{split}
        &\bar{M}_A(\mb{k})= S(3J_1 + 2 \Delta) + 2 DS \sum_i\sin{\mb{k} \cdot \boldsymbol{\gamma}_i} + g\mu_B B_0,\\
        &\bar{M}_B(\mb{k})= S(3J_1 + 2 \Delta) + 2 DS \sum_i\sin{\mb{k} \cdot \boldsymbol{\gamma}_i} - g\mu_B B_0,
        \end{split}\\
        \begin{split}
        &\bar{f}_1(\mb{k})= \frac{\mathcal{K}S}{2}\left( e^{-i \mb{k} \cdot \boldsymbol{\delta}_3}- e^{-i \mb{k} \cdot \boldsymbol{\delta}_1}\right) + i \Gamma_{\scriptscriptstyle \mathcal{K}} S e^{-i \mb{k} \cdot \boldsymbol{\delta}_2}
        + i \Gamma_{\scriptscriptstyle \mathcal{K}}^\prime S \left( e^{-i \mb{k} \cdot \boldsymbol{\delta}_3} + e^{-i \mb{k} \cdot \boldsymbol{\delta}_1}\right).\\
        & \bar{f}_2(\mb{k})= J_1 \sum_{i} \sin{\mb{k} \cdot \boldsymbol{\delta}_i} + \frac{\mathcal{K}S}{2}\left( e^{-i \mb{k} \cdot \boldsymbol{\delta}_3} + e^{-i \mb{k} \cdot \boldsymbol{\delta}_1}\right).
    \end{split}
    \end{align}
\end{subequations}
\endgroup
Here, $\mathcal{H}_N(\mathbf{k})$ also satisfies a generalized eigenvalue problem, and in this context the metric matrix is given by $\eta = \mathrm{diag}(1,1,-1,-1)$. $\boldsymbol{\delta}_i$ and $\boldsymbol{\gamma}_i$ denote the vectors connecting the nearest and next nearest neighbour sites.
\end{widetext}
\bibliography{main}

\end{document}